\documentclass{article}

\usepackage{arxiv}

\usepackage[utf8]{inputenc} % allow utf-8 input
\usepackage[T1]{fontenc}    % use 8-bit T1 fonts
\usepackage{hyperref}       % hyperlinks
\usepackage{url}            % simple URL typesetting
\usepackage{booktabs}       % professional-quality tables
\usepackage{amsfonts}       % blackboard math symbols
\usepackage{nicefrac}       % compact symbols for 1/2, etc.
\usepackage{microtype}      % microtypography
\usepackage{lipsum}		% Can be removed after putting your text content
\usepackage{graphicx}
\usepackage{natbib}
\usepackage{doi}
\usepackage{subfigure}
\usepackage{caption}
\usepackage{bm}
\usepackage{multirow}
\usepackage{threeparttable}
\usepackage{amsmath,amssymb}
\usepackage{enumerate,enumitem}
\usepackage{relsize}
\usepackage{soul}
\usepackage{adjustbox}

\DeclareMathOperator{\dist}{dist}
\DeclareMathOperator{\coun}{count}
\DeclareMathOperator{\infl}{influ}
\title{Robust Privacy-Preserving Motion Detection and Object Tracking in Encrypted Streaming Video}

\date{July 24, 2020}	% Here you can change the date presented in the paper title
\author{
	{Xianhao~Tian} \\
	Shenzhen University\\
	\texttt{tianxianhao2019@email.szu.edu.cn} \\
	\And
	{Peijia~Zheng} \\
	Sun Yat-Sen University\\
	\texttt{zhpj@mail.sysu.edu.cn} \\
	\And
	{Jiwu~Huang} \\
	Sun Yat-Sen University\\
	\texttt{jwhuang@szu.edu.cn} \\
}

\renewcommand{\shorttitle}{\textit{arXiv} Template}

\hypersetup{
pdftitle={Robust Privacy-Preserving Motion Detection and Object Tracking in Encrypted Streaming Video},
pdfsubject={eess.IV},
pdfauthor={Xianhao~Tian, Peijia~Zheng, Jiwu~Huang},
pdfkeywords={Encrypted video processing,cloud computing,video surveillance,motion detection,object tracking,compressed-domain feature},
}

\begin{document}
\maketitle

\begin{abstract}
Video privacy leakage is becoming an increasingly severe public problem, especially in cloud-based video surveillance systems.
It leads to the new need for secure cloud-based video applications, where the video is encrypted for privacy protection.
Despite some methods that have been proposed for encrypted video moving object detection and tracking,
none has robust performance against complex and dynamic scenes.
In this paper, we propose an efficient and robust privacy-preserving motion detection and multiple object tracking scheme for encrypted surveillance video bitstreams.
By analyzing the properties of the video codec
and format-compliant encryption schemes, we propose a new compressed-domain feature to capture motion information in complex surveillance scenarios.
Based on this feature, we design an adaptive clustering algorithm for moving object segmentation
with an accuracy of 4$\times$4 pixels.
We then propose a multiple object tracking scheme that uses Kalman filter estimation and adaptive measurement refinement.
The proposed scheme does not require video decryption or full decompression and has a very low computation load.
The experimental results demonstrate that our scheme achieves the best detection and tracking performance compared with existing works in the encrypted and compressed domain.
Our scheme can be effectively used in complex surveillance scenarios with different challenges,
such as camera movement/jitter, dynamic background, and shadows.
\end{abstract}

% keywords can be removed
\keywords{Encrypted video processing\and
	cloud computing\and
	video surveillance\and
	motion detection\and
	object tracking\and
	compressed-domain feature}

\section{Introduction}
\begin{figure}
	\centering
	\includegraphics[width=3.5in]{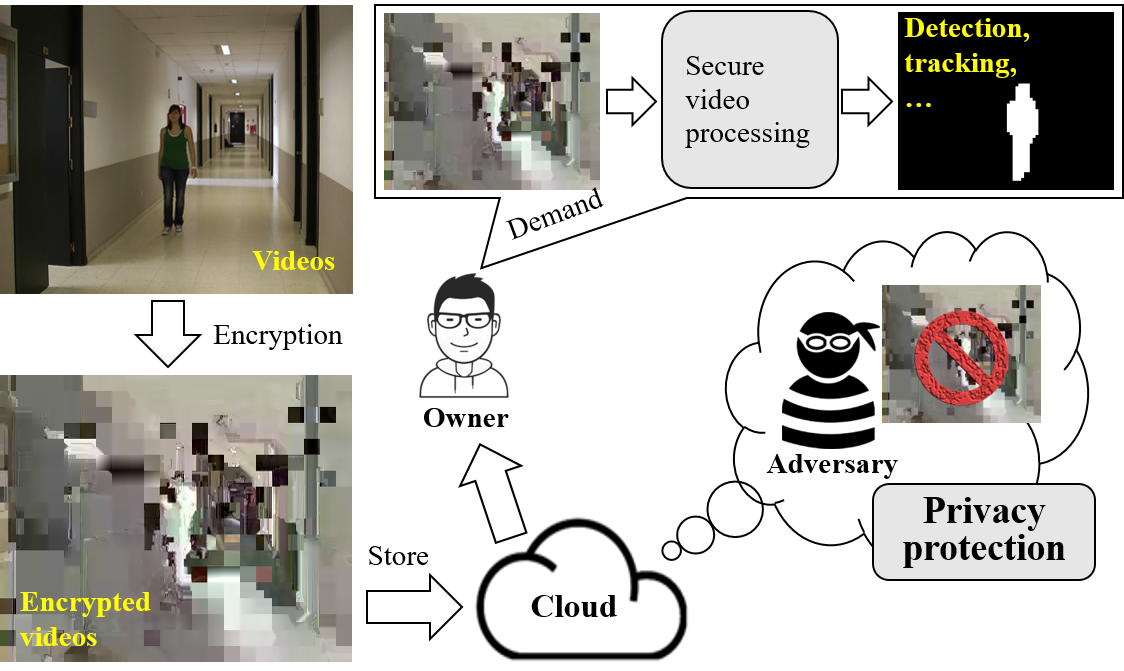}
	\caption{Illustration of the problem.}
	\label{Problem}
\end{figure}

Surveillance cameras play a significant role in public security.
The resulting surveillance videos are usually compressed into bitstreams for efficient transmission and storage.
Due to the heavy burden of storing and processing massive video data,
the video owner will prefer to outsource the expensive data storage and video signal processing tasks to the cloud, so as to enjoy conveniently personalized computing services and easy remote access from phones or PCs.
However, storing unencrypted data in the cloud
can threaten the privacy of the people that have been recorded.
The consequences can range from family privacy infringement to the leakage of secret information.
To avoid the exposure of video content, we must encrypt video bitstreams before storing them in the cloud server.
However, the encryption of video bitstream data poses a dilemma for the cloud, making subsequent video processing more complicated and difficult.
Therefore, it is urgently necessary to develop a secure cloud-based solution that can directly analyze and process an encrypted video stream in real time, so that applications such as anomaly alarms, abnormal object detection, and visual tracking can operate as usual in intelligent monitoring systems.
We illustrate this critical problem in Fig. \ref{Problem}.

Signal processing in the encrypted domain (SPED) \citep{6375935}, which enables the processing of ciphertext data without decryption, is a promising solution for privacy-preserving cloud computing.
Numerous SPED-based cloud computing applications have recently been proposed by using homomorphic encryption (HE) \citep{10.1007/3-540-48910-X_16,10.1145/2633600}.
% reviewer2, comment4 增加同态加密应用例子

To reduce ciphertext expansion, an image homomorphic encryption scheme~\citep{zheng2013efficient} is proposed. %using HE.
In~\citep{10.1007/978-3-642-03168-7_14}, Erkin \emph{et al}. presented a privacy-enhanced face recognition system based on the Paillier encryption. Troncoso-Pastoriza \emph{et al}.~\citep{6514926} proposed a private face verification system by employing quasi-fully HE.
Hu \emph{et al}.~\citep{7469805} proposed an outsourced feature extraction protocol on encrypted images using somewhat HE.
In \citep{10.1145/2886777}, Hu \emph{et al}. presented a secure nonlocal denoising algorithm on encrypted images with the Paillier encryption. Some reversible data hiding schemes \citep{8013725,7107988,8700282} are proposed on encrypted images using HE.
However, this HE-based approach is inefficient and impractical for real-time video encryption
due to the formidable size of the computations and the huge storage requirements.
For real-time encrypted video surveillance, one essential requirement is to utilize an efficient video bitstream encryption scheme to ensure that the required processing can be directly performed on the encrypted bitstream. In the following, we will mainly focus on the format-compliant video encryption,
which encrypts only certain specific syntax elements in the bitstream,
thus preserving the original video format and allowing the encrypted bitstream to be correctly parsed by common video decoders.

Motion detection and tracking are the key components of intelligent surveillance systems; these processes involve locating the positions of moving objects and tracking them over time in every frame of a surveillance video. Existing video motion detection and tracking schemes in the plaintext domain can be classified into two categories based on the methods used: the first is performed in the pixel domain, and extracts features from visual data using techniques such as optical flow \citep{duncan1992detection}, background modeling/subtraction \citep{784637,6238922,6975239}, and convolutional neural networks \citep{7485869,Redmon_2016_CVPR},
while the second operates in the compressed domain, and relies on syntax elements extracted from the video bitstream, such as macroblock coding bits, macroblock partition, motion vectors, quantization parameters and discrete cosine transform (DCT) coefficients \citep{6272352,6490026,7801078}. The first type of approach is computationally expensive and often has bottlenecks in real-time operations due to the high complexity of these algorithms and the enormous amounts of data involved in processing.
Moreover, without fully video decompression, pixel-domain approaches cannot be directly performed on cloud surveillance video data stored in compressed form.
In contrast, compressed-domain approaches can run at high speeds, since they require only partial decoding of the bitstream.
Therefore, compressed-domain approaches are preferable for secure real-time video surveillance in cloud computing.

There are already some privacy-preserving video surveillance schemes
relying on pixel-domain motion detection techniques,
where the encryption methods are applied to the decompressed video frames. Upmanyu \emph{et al}. \citep{5459370} designed a privacy-preserving motion detection and tracking system for encrypted surveillance video by using a secret sharing scheme with the Chinese remainder theorem. In \citep{zeng2010object}, Zeng \emph{et al}. proposed a secure object detection scheme based on Paillier encryption. Chu \emph{et al}. \citep{10.1145/2502081.2502157} presented a privacy-preserving moving object detection scheme that scrambles each frame with matrix permutation and multiplication. In \citep{10.1007/978-3-319-27671-7_47}, Jin \emph{et al}. proposed a video foreground extraction method based on chaotic-mapping video encryption.
In \citep{lin2017moving}, Lin \emph{et al}. proposed a system that separated video pixels into two parts, one of which was kept unchanged and the other encrypted with pixel scrambling. Since the resultant encrypted video are encrypted frame sequences in~\citep{10.1145/2502081.2502157,10.1007/978-3-319-27671-7_47,lin2017moving}, the adaptive Gaussian mixture models (GMM) \citep{784637} can be applied to obtain the motion detection results.
However, frame-encryption based schemes usually result in high execution times and also have apparent drawbacks in terms of transmission and storage because they do not support video compression.

Following video compression standards and format-compliant video encryption techniques~\citep{5955103},
some recent works enable motion detection and tracking on encrypted video bitstreams.
Guo \emph{et al}. \citep{10.1145/3131342} designed a motion detection and tracking scheme for an H.264/AVC~\citep{1218189} bitstream encrypted with the format-compliant encryption proposed in \citep{6725633}.
They proposed a feature based on the length of motion vector difference (LMVD),
and a region update algorithm for object tracking.
In \citep{8451279}, Ma \emph{et al}. used the format-compliant encryption scheme proposed in \citep{6589970} to encrypt an H.265/HEVC~\citep{6316136}
video, and then shuffle the coding tree units (CTU).
To perform motion detection and tracking on encrypted H.265 bitstreams,
they proposed to extract features from coding bits (CB) and partition depth (PD).
These bitstream-based schemes have the advantages of low storage requirements and high computational efficiency.
However, these schemes can only detect objects in 16$\times$16 block level or CTU level (between 16$\times$16 pixels and 64$\times$64 pixels).
Moreover, they do not have robust detection and tracking performance in complex surveillance scenarios presenting different challenges, such as camera shaking/moving, shadow, dynamic background, etc.

In this paper, we focus on secure motion detection and tracking in the cloud-based video surveillance system,
where surveillance video is compressed and encrypted with format-compliant video encryption.
Without loss of generality, we use H.264/AVC as an example to illustrate our scheme, considering its popularity in video surveillance compression system. We propose a privacy-preserving motion detection and tracking scheme to be applied to encrypted video bitstream in the cloud.
More specifically, we propose a novel compressed-domain feature known as the density of non-zero residual coefficient (DNRC)
to extract motion information from the encrypted video bitstream. The proposed feature enables us to successfully locate smaller objects than the existing features used in \citep{10.1145/3131342,8451279}.
To efficiently separate motion regions from the background,
we design an adaptive and non-parameterized clustering method based on the density-based spatial clustering of applications with noise (DBSCAN) \citep{ester1996density}.
We then propose a reliable object tracking scheme with Kalman filter estimation and an adaptive bounding box refinement method against different challenges in real surveillance tasks.
The major contributions in this paper can be summarized as follows.
\begin{itemize}
	\item We propose a novel feature, DNRC, extracted from encrypted video bitstreams.
	The proposed feature locates objects with an accuracy of 4$\times$4 pixels,
	which is more precise than the 16$\times$16 block-level of previous works in the encrypted and compressed domain.
	
	\item We design an adaptive and non-parameterized clustering algorithm based on DBSCAN,
	which can effectively segment moving objects from the dynamic background.
	
	\item We propose a reliable multi-object tracking scheme based on Kalman filtering by adaptively refining the observation data.
	Our method has robust performance in complex scenarios presenting different challenges, such as occlusion, camera moving/jittering, shadow.
	
	\item Our secure motion detection and tracking scheme allows for effective detection, robust tracking,
	and fast computation.
	The experimental results show that the proposed algorithms achieve the best performance among comparable schemes for encrypted bitstream video.
\end{itemize}

The rest of this paper is organized as follows. Section \ref{statement} presents our threat model and encryption scheme.
In Section \ref{algorithm}, we introduce the feature DNRC, the proposed clustering algorithm, and our motion tracking scheme.
The experimental results conducted on different surveillance video databases are presented in Section \ref{experiment}, and some discussions are provided in Section \ref{discussion}.
Finally, we draw conclusions in Section \ref{conclusion}.

\section{Problem Statement}\label{statement}%Preliminary
\subsection{Threat Model}
\begin{figure}
	\centering
	\includegraphics[width=3.5in]{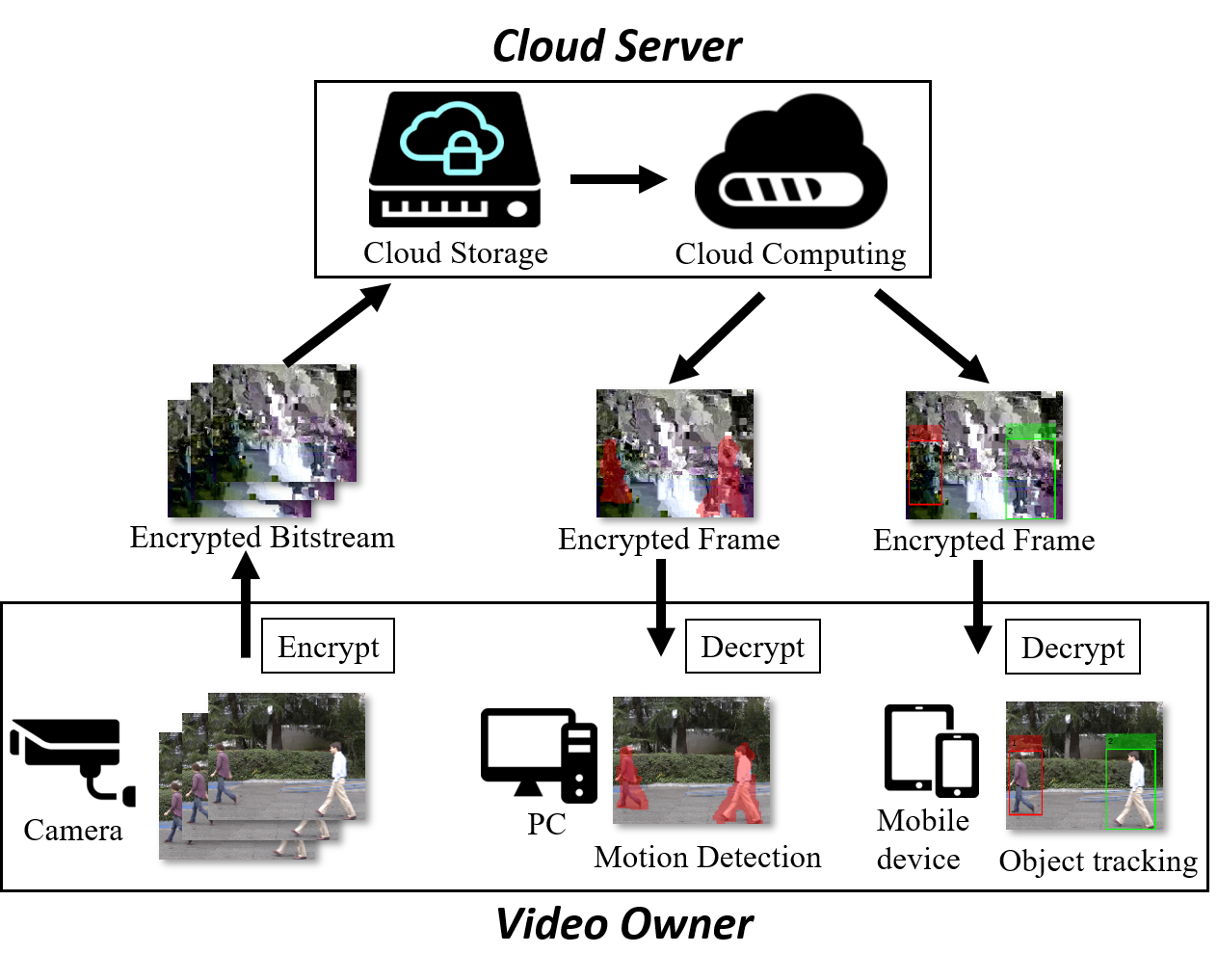}
	\caption{System model.}
	\label{systemmodel}
\end{figure}
In this paper, we consider a non-interactive video motion detection and tracking system that involves only two parties: the video data owner and the cloud server.
The owner uses several devices simultaneously, such as surveillance cameras, computers, smartphones, tablets, etc.
A sketch of our system model is shown in Fig.~\ref{systemmodel}.
%放到段末会好一点。句子读起来顺畅。突然插个图进来对下面句子描述没辅助作用。
We assume that the owner's storage and computation resources are limited, and that in order to deal with a vast volume of video data, the owner needs to outsource data storage and computing tasks to the cloud.
%\textcolor{red}{A sketch of our system model is shown in Fig.~\ref{systemmodel}. The cloud server can only access the encrypted data of video owner.}

Following most existing privacy-preserving multimedia outsourcing schemes~\citep{6216412,7469805},
we adopt a semi-honest setting throughout this paper.
Specifically, the cloud server is considered a semi-trusted adversary in our application scenario,
which honestly executes the proposed protocol but attempts to learn additional sensitive video content, such as human faces and card security codes, from the encrypted video and all the exchanged messages.
To avoid the leakage of private or sensitive content to the cloud, the owner prefers to store encrypted video on the cloud storage server.
In this paper, we use format-compliant selective encryption, a widespread technique in multimedia encryption, due to its advantages in maintaining a balance between privacy and convenience.
More detailed discussions of privacy protection with selective video encryption can be found in
\citep{5733402}.
%Section \ref{securediscuss}.
By using our protocols, the cloud server can detect moving objects and obtain object trajectories. However, without the decryption keys, the cloud server cannot deduce sensitive and private information about the individuals in the video.
%\st{, e.g., exact human faces or card security codes}
%上文已经强调过了private information的内容了。重复说明可删除。

\subsection{Video Encryption Scheme}
In our video encryption scheme, we encrypt the codewords of three syntax elements, i.e., motion vector difference (MVD), residual data, and intra-prediction mode (IPM).
Each syntax element is encrypted according to its specified coding method. %The details of encryption on different syntaxes are given as follows.

\subsubsection{MVD Encryption}
MVDs are encoded with the $k$-order Exp-Golomb (EG$k$) code in H.264/AVC. For example, the codeword structure of EG0 ($k$ = 0) can be represented as $$[m\ \text{zeros}, 1, \text{suffix}],$$ where ``$m$ zeros'' indicates the length of ``suffix'', and ``suffix'' contains $m$-bits of the encoded information. We encrypt the last bit of the non-zero MVD codeword. The last bit encryption may change the sign of MVD and satisfies the format compliance without bit increase.
\subsubsection{Residual Data Encryption}
The residual data are obtained from the intra or inter predictions. After transformation and quan-tization, the data are entropy encoded using CAVLC in the baseline profile. Using CAVLC, each codeword of the residual data can be represented as
$$\begin{aligned}
&[\emph{coeff\_token}, \emph{sign\_of\_trailingones}, \emph{level}, \emph{total\_zeros}, \emph{run\_before}].
\end{aligned}$$
The codeword of \emph{level} can be further represented as $$ [\emph{level\_prefix}, \emph{level\_suffix}]. $$ We encrypt all bits in the codeword of \emph{sign\_of\_traillingones} and the last bits in the codeword of \emph{level}. For format compliance,  \emph{level} cannot be encrypted when its suffix length is equal to 0.
\subsubsection{IPM Encryption}
In the \emph{high profile} of H.264, three types of intra coding are supported to be encrypted, which are denoted as Intra\_4$\times$4, Intra\_8$\times$8, and Intra\_16$\times$16. The IPM for Intra\_16$\times$16 blocks are specified in the
\emph{mb\_type} (macroblock type), which defines other block parameters like coded block pattern (CBP). The \emph{mb\_type} is encoded with the Exp-Golomb code. To ensure format compliance of the encrypted bitstream, we cannot change the CBPs of \emph{mb\_type} when encrypting IPM. According to the \emph{mb\_type} table for the I frame \citep{telecom2003advanced}, only the last bit of the codewords of all \emph{mb\_type} of Intra\_16$\times$16 can be encrypted in the I frame. However, this encryption method cannot be applied to those Intra\_16$\times$16 blocks in the P frame due to the different \emph{mb\_type} tables used in the I and P frames \citep{telecom2003advanced}. To keep the length of the codeword unchanged and format-compliant, we only encrypted the last bit of those codewords of \emph{mb\_type}s in the set of
$$\{7,8,11,12,15,16,19,20,23,24,27,28\}.$$
The Intra\_4$\times$4 and Intra\_8$\times$8 blocks use the same method to code their IPMs. The H.264 codec provides every 4$\times$4/8$\times$8 block with a most probable intra prediction model (MP-IPM) based on the neighbor blocks. If the MP-IPM is equal to the lowest-bit-cost intra prediction model, then the codeword of IPM only consists of one flag bit ``1''. Otherwise, the codeword is composed of one flag bit ``0'' and three bits fixed-length code indicating one of the other eight modes except MP-IPM. These three bits can be encrypted without format conflict concern. Generally, our encryption scheme encrypts all IPMs in the Intra\_4$\times$4 block and Intra\_8$\times$8 for the I and P frames.
The target bits for encryption are first extracted from the video bitstream and then encrypted by a standard cipher, such as RC6 block cipher.
The enciphered bits are placed back into the bitstream to replace the original bits.
During video decryption, the enciphered bits in the encrypted video bitstream are decrypted to the plaintext bits.
We use these plaintext bits to replace the corresponding enciphered bits in the video bitstream.
Note that the proposed technique for object tracking is not designed to rely on this specific video encryption scheme.
We will discuss the extension to other video coding standards and video
bitstream encryption schemes in Section \ref{adaptation}.

\begin{figure}[!t]
	\centering
	\includegraphics[width=2.8 in]{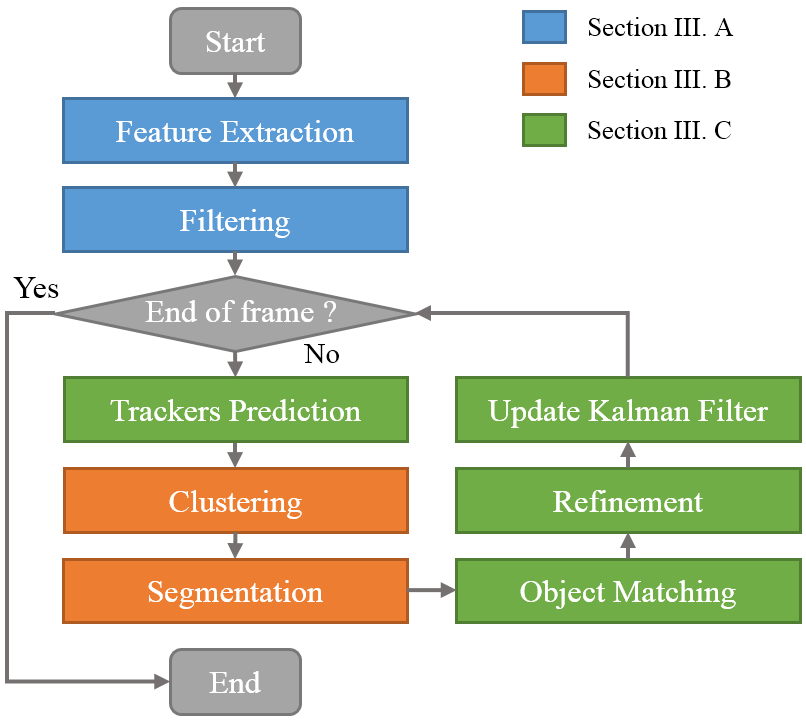}
	\caption{The flowchart of our motion detection and tracking scheme.}
	\label{algorithm flowchart}
\end{figure}

\section{Proposed Privacy-Preserving Motion Detection and Tracking Scheme}\label{algorithm}
We summarize our scheme's steps in Fig. \ref{algorithm flowchart}, which includes feature extraction, clustering for initial motion detection, and object tracking and refinement.
We extract the proposed feature in the encrypted bitstream without fully video decoding and design a temporal filter for smoothing (Section III-A).
The motion detection and tracking procedure is implemented frame by frame.
Based on the extracted features, we propose a clustering algorithm to perform segmentation and obtain initial detection results (Section III-B).
We then propose a refinement strategy and adopt adaptive Kalman filtering to obtain precise detection results and track moving objects (Section III-C).

\subsection{Feature Extraction}\label{extraction}
In the video encoding process, the macroblocks in a motion region always contain considerable variation in pixel values and many detail textures,
When performing DCT on these macroblocks, a large number of residual coefficients are generated.
%意思表达改进一下会更好
%\textcolor{red}{which result complex residual data. Therefore, these macroblock has more non-zero residual coeffcients after DCT transform.}
%\st{Although the codewords of residual coefficients are encrypted in most format-compliant video encryption schemes, the statistical characteristics of the quantity of the residual coefficients are still preserved in the encrypted bitstream to ensure that the compressed video bitstream can be decoded correctly.}
%放到段末统一说明。根据reviewer1 对加密算法的特征保留进行阐述。
There are two kinds of methods modifying the residual coefficients in existing format-compliant H.264 video encryption schemes, i.e., sign inversion and value modification on residual coefficients.
Both the sign inversion and value modification do not change the number of non-zero residual coefficients in a luma block.
Therefore, although the codewords of residual coefficients are encrypted in format-compliant video encryption schemes,
the number of non-zero residual coefficients in every frame is preserved after video encryption.
Some reports \citep{6490026,POPPE2009428,6151836} have shown that the %\st{prediction}
residuals are capable of detecting moving objects in the plaintext domain.
Non-zero residual coefficients are more likely to be concentrated in the motion region, meaning that a block with more non-zero residual coefficients has a higher probability of being in the moving object region.
%\textcolor{red}{According to a recent survey on H.264 encryption algorithm \citep{TABASH201920}, to ensure that the encrypted video bistream can be decoded correctly, two possible encryption processing can be applied to residual coefficients: random sign inversion and random value permutation. However, both of them have no impact on  the quantity of non-zero residual coefficients. That is to say, DNRC is preserved after format-compliant encryption. }

We divide the $k$-th frame into non-overlapping 4$\times$4 blocks $\{{u}_{lm(k)}\}_{lm} \triangleq U(k)$.
We can also use the sequence $\{ u_i(k) \}_i$ to
represent the 2D matrix $U(k)$ in a raster scan order.
Without causing confusion, we sometimes use $u_i$ and $U$ in the following.
According to the H.264/AVC compression standard, such a block ${u}_{i}$ is either located in a whole 4$\times$4 transform block or combined with three more blocks to form an 8$\times$8 transform block.
Based on the above analysis, we propose a new feature, i.e., DNRC, in each 4$\times$4 block ${u}_{i}$.
The proposed feature approximately assesses the density of non-zero residual coefficients within a 4$\times$4 block.
We use $\coun ( \cdot)$ to count the number of non-zero residual coefficients in a 4$\times$4 or 8$\times$8 residual coefficient block.

We firstly consider the DNRC feature extraction in P frames.
When using Intra\_16$\times$16 prediction mode,
the DC coefficients of each 4$\times$4 block in the Intra\_16$\times$16 macroblock are separated from the residual coefficient block to form another block, on which a Hadamard transform is performed.
%\textcolor{red}{all DC residual coefficients in a macroblock will be extracted to form a new 4$\times$4 matrix and then undergo a second 4$\times$4 transform. If a 4$\times$4 DCT block has a non-zero DC residual coefficient that is decoded from the DC block, its DNRC should increase 1.}
The total number of non-zero residual coefficients for ${u}_{i}$ then should increase by one if ${u}_{i}$ has a non-zero DC component.
Hence,
we compute the influence of the DC component on the DNRC of ${u}_{i}$ as
\begin{equation}
\infl\left ( {u}_{i}\right ) =\left\{\begin{array}{ll}{1,} & {\text { if } {u}_{i} \text { has non-zero DC component }} \\ {0.} & {\text { otherwise }}\end{array}\right.
\end{equation}
The DNRC of a 4$\times$4 block $u_{i}$ is then defined as
\begin{equation}
{d}_{i}= \coun \left ( {u}_{i}\right )+ \infl \left ({u}_{i}\right ).
\end{equation}

\begin{figure}[!t]
	\centering
	\includegraphics[width=2.5in]{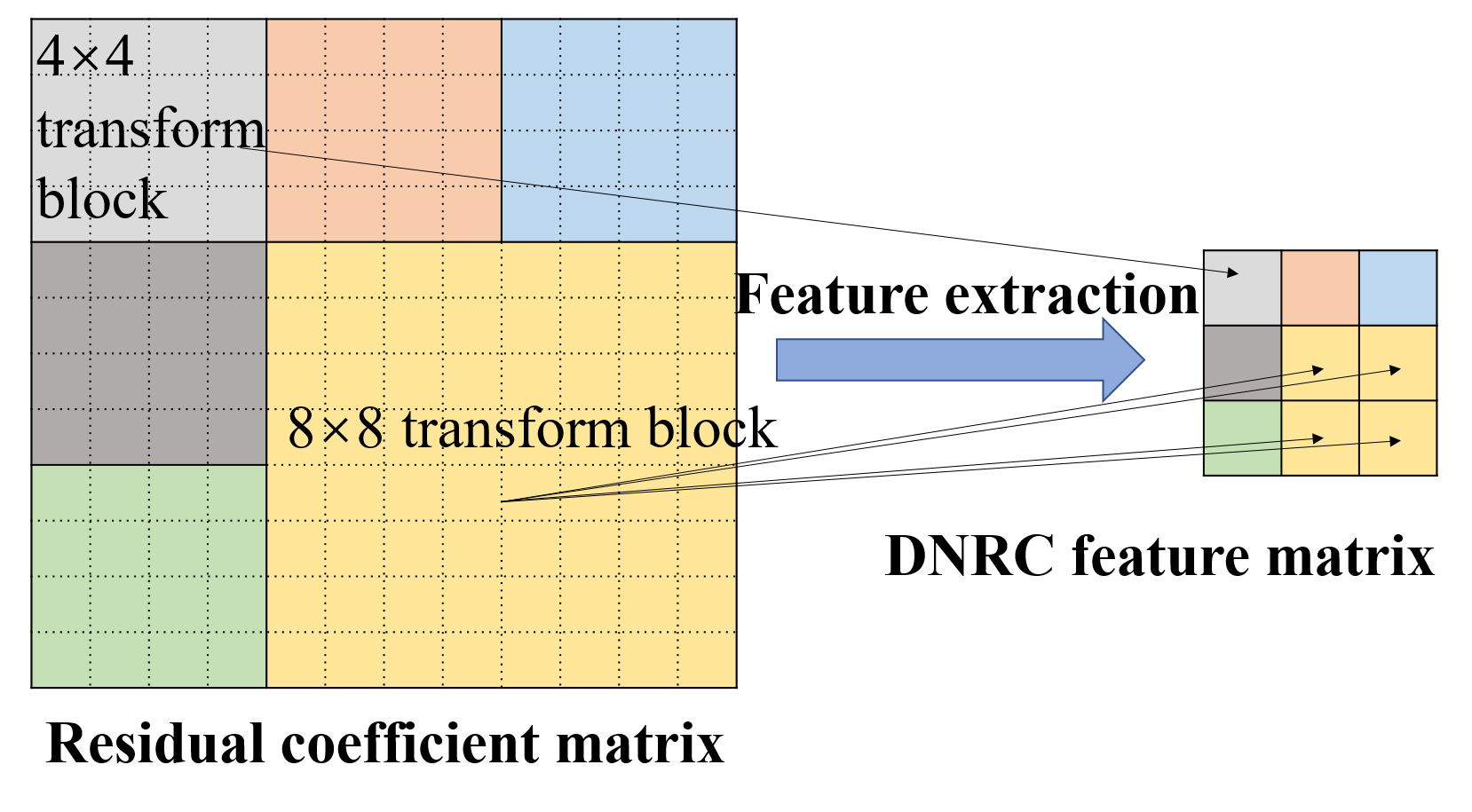}
	\caption{Some example DNRC extraction in 4$\times$4 and 8$\times$8 transform blocks.
		The left and right matrices denote the residual coefficient matrix and the DNRC feature matrix, respectively.}
	\label{fig:featureExtr}
\end{figure}

When the H.264/AVC encoder additionally uses
%allows using
a transform block size of 8$\times$8,
it will produce an 8$\times$8 residual coefficient block covering four 4$\times$4 ${u}_{i}$s,
rather than four residual coefficient blocks size of 4$\times$4.
In order to extract feature information from such a ${u}_{i}$ in this case, we set the DNRC of ${u}_{i}$ as the number of non-zero residual coefficients in the 8$\times$8 residual coefficient block.
Considering that there are only 16(=4$\times$4) elements in ${u}_{i}$, we constrain its maximum DNRC to be 16.
Thus, for a 4$\times$4 block ${u}_{i}$ located in an 8$\times$8 transform block ${\hat{u}}_{i}$,
its DNRC value is computed as
\begin{equation}
{d}_{i}=\min   ( \coun  ( {\hat{u}}_{i}   ) , 16   )
\end{equation}
where $\min(\cdot)$ denotes the minimum operator.
We illustrate some examples of DNRC extraction in 4$\times$4 and 8$\times$8 transform blocks in Fig.~\ref{fig:featureExtr}.

Since the H.264/AVC coder generates the prediction residual in the I frame by referring to the neighboring macroblock in the spatial domain, the residual data in the I frame
%过于冗长。一句话表达清晰。
%\textcolor{red}{Since the I frame only use intra prediction, its residual data}
cannot represent the motion information in the temporal domain.
%\st{in the temporal domain} .
Thus, we compute the DNRC value in the I frame from the DNRC values in the neighboring P frames using the inverse distance weighted interpolation method \citep{RN79}.
Suppose that ${d}_{i}(k)$ is the DNRC of the block ${u}_{i}$ in the $k$-th frame.
We initialize ${d}_{i}(k)$s in the I- frames as zeros, and then update ${d}_{i}(k)$ in the I-frame as
\begin{align}
d_{i}(k)=\sum_{\tau=k-\delta}^{k+\delta} \frac{d_{i}(\tau)}{|k-\tau|} / \sum_{\substack{\tau=k-\delta \\ \tau \neq k}}^{k+\delta} \frac{1}{|k-\tau|}
\end{align}
where $\delta$ denotes the window size of the interpolation.

Although our DNRC feature can reflect the motion information, other pixel changes in the video content, including the foreground and background pixel changes, will also influence the DNRC feature.
These DNRC values will create false-positive motion regions.
For the DNRC values indicating real moving objects, we consider these false-positive values as ``noises'' in the feature sequence $\{d_i(0), d_i(1), \cdots \}$.
%In general, the time sequence $\{d_i(0), d_i(1), \cdots \}$ contains several noise values, which
These noise values are likely to be caused by dynamic background noise or compression quantizing noise, such as valley noise (a zero DNRC surrounded by consecutive non-zero DNRCs in the form of a valley) or spike noise (a short-term, non-zero DNRC sequence that appears abruptly in a zero sequence).
For valley noise in the $k$-th frame, we use the smallest DNRC in the ($k$-1)-th and ($k$+1)-th frames to replace the zero DNRC in the valley noise, while spike noise can be diminished by applying a lowpass filter along the temporal direction.
Due to the sensitivity of DNRC between zero and non-zero,
we also want to set zero the noisy DNRC values that need to be filtered out,
and keep the normal DNRC values unchanged.
%根据审稿人1的comment10，噪声可能表达不是很好，我们可以重新描述一下这一段。
%\textcolor{red}{Inevitably, the residual coefficients that belong to background noises also will be extracted. Background noise, such as swaying tree branches and moving shadows, causes drastic pixel changes and therefore brings large noise DNRC. But it is often very transient. For a background block ${u}_{i}$, this noise DNRC looks like a spike in the time sequence $\{d_i(0), d_i(1), \cdots \}$. We design a special temporal filter to filter out the ``spike'' noise. Opposite to the spike noise, there is a kind of noise that looks like a narrow valley in the time sequence. It is because the pixels in the object center region that has the same color keep unchanged. However, the missing DNRC in the ``valley'' can be restored by temporal filtering.}
Assume that the distribution of noise spikes is sparse, i.e., there is one noise spike approximately every $\mu$ frames.
We use $\mathbb{K}$ to denote
a sub-set of frame indexes,
% V的真正意思是valley noise的集合, reviewer 2 的comment 5.
%\textcolor{red}{the collection of frame indexs of ``valley'' noise for block ${u}_{i}$,}
i.e.,
\begin{align}
\mathbb{K} &= \bigg \{ {k} \, \bigg| {d}_{i}(k)=0 \mbox{ and } \prod_{\substack{\tau=k-\delta \\ \tau \neq k} }^{k+\delta} {d}_{i}(\tau)>0 \bigg \},
\end{align}
The filtered DNRC value ${f_i}(k)$ can then be obtained as
\begin{align}
{f_i}(k)&=\left\{\begin{array}{ll}
{\min\left({{d}_{i}(k+1),{d}_{i}(k-1)}\right),} &{ k \in \mathbb{K}} \\
{{d}_{i}(k) \cdot \mathcal{Z} \left(\sum\limits_{\sigma=k-\mu}^{k} \prod\limits_{\tau=\sigma}^{\sigma+\mu} {d}_{i}(\tau) \right),} &{k \notin \mathbb{K}}
\end{array}\right.
\end{align}
where $\mathcal{Z}(\cdot)$ is a function deciding whether the input is zero or not, i.e.,
%\begin{align}
$\mathcal{Z}(x)=
\left\{
\begin{array}{ll}
{0,} &{ x = 0 } \\
{1,} &{ x \neq 0}
\end{array}
\right.
$.
%\end{align}
We denote all the filtered DNRC values in the $k$-th frame by the sequence ${\bf f}(k) = \{{f_0}(k),{f_1}(k),\cdots  \}$.
As shown in Fig.~\ref{fig:featureExtr},
we can reshape ${\bf f}(k)$ into a 2D matrix ${F}(k)$, called the feature image, according to the positions of $u_i(k)$ in $U(k)$.
An example of the feature image can be seen in Fig. \ref{detection}(c).
Without causing confusion, we will use $f_i$ and $F$ in the following.

\subsection{Clustering}\label{motion detection}
Based on the feature image $F$,
we propose a clustering method to segment initial motion regions from the background.
A large value of $f_{i} \in F$ will make a great contribution to gathering other non-zero elements in $F$ into a cluster.
In general, noise regions such as the shadows of moving objects usually have small DNRC values,
and disperse into several small areas that are difficult to form into a complete cluster.
In view of these characteristics, we propose an efficient and adaptive clustering algorithm based on DBSCAN, in which the number of clusters in a feature image need not be specified.

Let us denote the feature image consisting of all non-zero elements in $F$ by $F'=\{f_{i}\in F |  f_{i} > 0\}$.
Given a point $f_{i} \in F'$ and a radius $\epsilon>0$, we define the $\epsilon$-neighborhood as
\begin{equation}
\label{E1}
N_{\epsilon}\left(f_{i}\right)=\left\{f_{j} \in F'  |  \dist\left(f_{i}, f_{j}\right) \leq \epsilon\right\}
\end{equation}
where $\dist (f_{i}, f_{j} )$ is a pre-defined distance between $f_{i}$ and $f_{j}$.

If the point number of $N_{\epsilon} (f_{i} )$ is greater than a threshold minPts,
DBSCAN will consider $f_{i}$ a core point of a new cluster.
Traditionally, minPts is usually set to one more than the cluster's dimension,
e.g., minPts = 3 in our application scenario.
All the points in ${N}_{\epsilon} (f_{i} )$ will form a temporary cluster of $f_{i}$.
If a point $f_{j}$ in this temporary cluster is a core point, all the points in $N_{\epsilon} (f_{j} )$ will be included to form a new temporary cluster.
This process is repeated until no more new points are added to the temporary cluster, and
the final cluster of $f_{i}$ is obtained.
After that, all the clusters are generated.
The points that do not belong to any cluster are regarded as noise points.

\begin{figure}[!t]
	\centering
	\subfigure[Plaintext]{
		\includegraphics[width=1in]{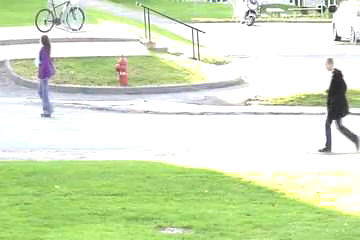}
	}
	\subfigure[Ciphertext]{
		\includegraphics[width=1in]{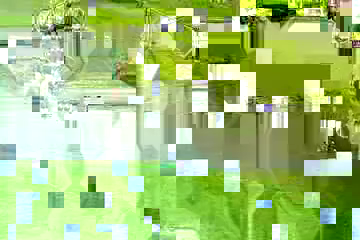}
	}
	\subfigure[Feature image]{
		\includegraphics[width=1in]{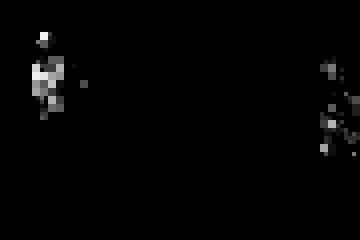}
	}
	\subfigure[Clustering]{
		\includegraphics[width=1in]{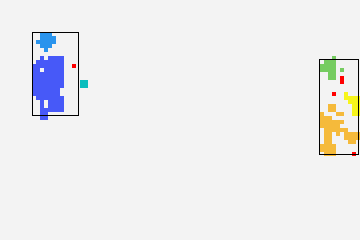}
	}
	\caption{Motion detection example using the 571th frame from \emph{pedestrians}. The red points in (d) indicate noise, and the different colors of other points indicate that they belong to different clusters. }
	\label{detection}
\end{figure}

\begin{figure}[!t]
	\centering
	\includegraphics[width=3.8in]{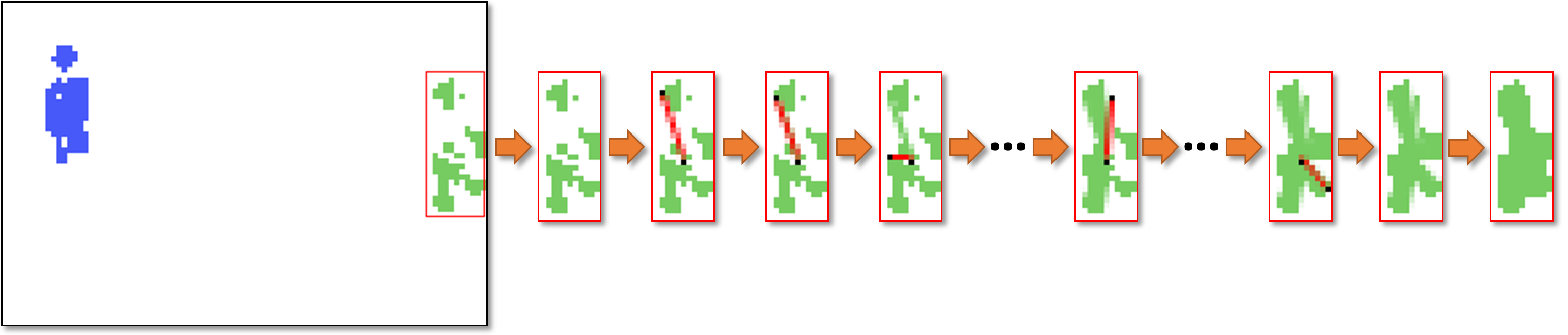}
	\caption{Example of how the blanks in the cluster of Fig. \ref{detection}(d) are filled in. First, we draw lines between the centroid and the other points in the cluster to connect each part, and then strengthen the filling effect using a 3$\times$3 morphological closing operation.}
	\label{FillBlank}
\end{figure}

The definition of the $\epsilon$-neighborhood plays a central role in the clustering algorithm.
To adapt to our application scenario, a point $f_{i}$ with a significant value should have a large radius $\epsilon_{i}$.
We must ensure that $\epsilon_{i} \geq \sqrt{2}$, since $\sqrt{2}$ is the minimum Euclidean radius of an 8-adjacent connection.
Thus, we set the radius of $f_{i}$ in our method as
\begin{equation}
\label{EE1}
\epsilon_{i}=\sqrt{2}+\eta \cdot {f}_{i}
\end{equation}
where $\eta$ is a factor representing the clustering ability of DNRC that needs to be defined in the following.

We use $\bar{f}$ to represent the mean of all $f_i$s in $F'$.
A higher value of $\bar{f}$ indicates that the number of non-zero residual coefficients generally is larger on every $4\times4$ pixel block.
It mainly results from fast-moving objects in the frame or suddenly shaking background caused by camera movement/jitter.
For a fast-moving object, clustering different parts into different small clusters with a small radius, respectively, can avoid involving too many false-positive regions.
As for camera movement/jitter, small clusters can separate the suddenly changed background from real motion regions.
Therefore, when the value of $\bar{f}$ is higher, using a smaller radius can improve the detection accuracy and decrease the false positive rate.
%添加的
Conversely, a smaller $\bar{f}$ indicates that the objects are moving slowly and the background is stable.
We can use a large radius to generate large-size clusters to avoid missing the false-negatives points without worrying about covering the background regions.
Thus, it is reasonable that $\eta$ is in inverse proportion to $\bar{f}$.

Specifically, we define $\eta$ as
\begin{equation}
\eta=\max \left(0, \frac{1}{2}\left(1-\log _{8} \bar{f}\right)\right)
\end{equation}
where
%\st{$\overline{f}$ is the mean value of all $f_i$s in $F'$, and}
$\max( \cdot )$ denotes the maximum operator.
We use $\gamma_{j}$ to evaluate the influence of the point $f_{j}$ on the core point $f_i$, i.e.,
\begin{equation}
\label{EE2}
\gamma_{j}=\eta \cdot {f}_{j}.
\end{equation}
%Assuming that $\dist_{E} (f_{i}, f_{j} )$ denotes the Euclidean distance between $f_{i}$ and $f_{j}$.
Consequently, we define the $\epsilon$-neighborhood of $f_{i}$ in the proposed method as
\begin{equation}
\label{E2}
N_{\epsilon}\left(f_{i}\right)=\left\{f_{j} \in F'  |  \dist_{E}\left(f_{i}, f_{j}\right) \leq \epsilon_{i}+\gamma_{j}\right\}
\end{equation}
where $\dist_{E} (f_{i}, f_{j} )$ denotes the Euclidean distance between $f_{i}$ and $f_{j}$.
By combining the inequality $\dist_{E}\left(f_{i}, f_{j}\right) \leq \epsilon_{i}+\gamma_{j}$ with Eq.~(\ref{EE1}) and Eq.~(\ref{EE2}), we get
\begin{equation}
\label{E8}
\dist_{E}\left(f_{i}, f_{j}\right) - \eta \cdot \left({f}_{i}+ {f}_{j}\right) \leq \sqrt{2}.
\end{equation}
We denote the terms on the left-hand of the above inequality by $g\left(f_{i}, f_{j}\right)$,
and rewrite Eq.~(\ref{E2}) as
\begin{equation}
\label{E3}
N_{\epsilon}\left(f_{i}\right)=\left\{f_{j} \in F' | g\left(f_{i}, f_{j}\right) \leq \sqrt{2}\right\}.
\end{equation}

By comparing Eq.~(\ref{E1}) and  Eq.~(\ref{E3}), we can see that we set
$\epsilon=\sqrt{2}$ and use $g\left(f_{i}, f_{j}\right)$ to adapt the DBSCAN algorithm to our application scenario.
Unlike the typical DBSCAN algorithm, we do not need to preset the radius $\epsilon$.
Thus, our clustering algorithm is adaptive and non-parameterized.
We show an example of the clustering results in Fig.\ref{detection}(d).

The proposed clustering method mainly relies on spatial information, and may not perform satisfactorily when moving objects have less detailed texture areas.
%承上启下，这里要清楚的就是valley noise。
%\textcolor{red}{The ``valley'' noises we mentioned in the temporal filtering of DNRC often generate a blank hole in the center area of motion region. }
For example, as shown in \ref{detection}(d), the blocks in the central area of the man on the right
have less detailed areas, because he is wearing entirely black clothing.
As a result, his body is separated into three clusters with the clustering method.
To solve this problem, we employ additional temporal
%\st{temporal}
%\textcolor{red}{tracker prediction}
information to improve the clustering results.
Suppose that $ {B}_{t,k-1} ^{l}$ is
the final bounding box of the $l$-th moving object in the ($k$-1)-th frame outputted by our object tracking algorithm in Section.~\ref{motion tracking}, which uses temporal information over the feature image sequence for target tracking.
In the $k$-th frame, if more than one cluster appears in any $ {B}_{t,k-1}^{l}$,
we will merge these clusters into a single cluster by filling in the blanks to produce a convex shape,
an example of which is given in Fig. \ref{FillBlank}.
For every cluster, we generate an initial bounding box to exactly contain its region, i.e., $ {B}_{i,k}^{l}$ for the $l$-th cluster in the $k$-th frame.

%\textcolor[rgb]{1.00,0.00,0.00}{In contrast, when multiple objects are obscured by each other, their clusters will be connected together.
%We therefore need to assume that the velocities and the sizes of the bounding boxes of the occluded objects remain unchanged during the occlusion.
%Based on this assumption, we adopt an approach similar to that in \citep{guo2017efficient} to solve the occlusion problem.
%Since clusters with small areas are likely to be background noise, we remove them using a minimum area filter, where the threshold of the filter is denoted as $\alpha$.
%}
%
%\textcolor[rgb]{1.00,0.00,0.00}{We then connect each fraction of the clusters by filling in the blanks to produce a convex shape.
%An example that demonstrates this filling process is given in Fig. \ref{FillBlank}.
%Finally, we magnify the feature image by four times to ensure it has the same size as the original frame.}

\begin{figure}[!t]
	\centering	
	\includegraphics[width=3in]{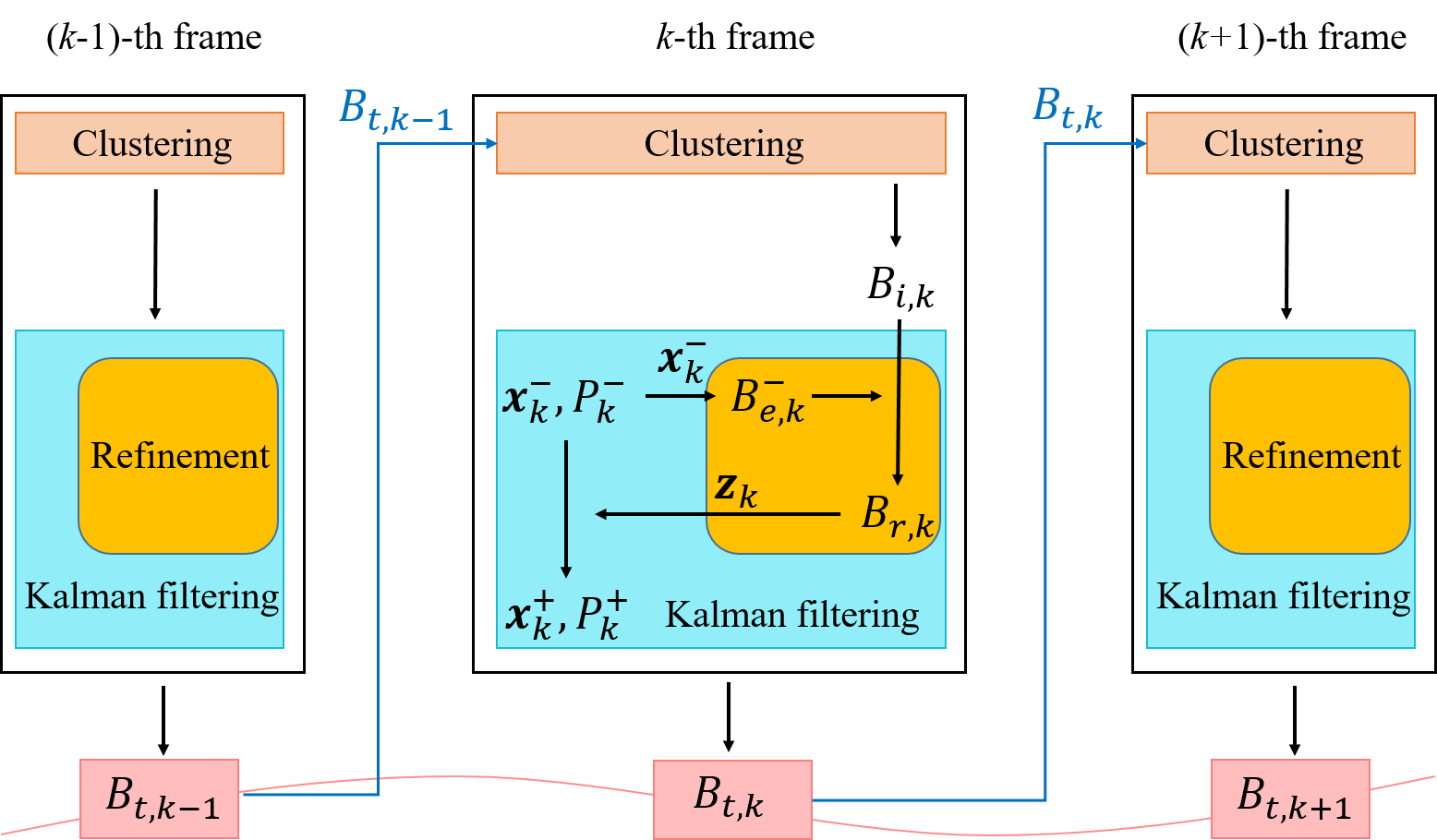}
	\caption{Diagram of the proposed tracking method}
	\label{fig:sketch}
\end{figure}

\subsection{Object Tracking}\label{motion tracking}
%
%\st{Since the employment of video encryption will noise the frame data, an isolated DNRC feature in one frame is limited to contain enough motion information.The clustering algorithm mainly relies on the isolated frame features, resulting in that the initial motion boxes $\{ {B}_{i,t}^{l} \}_l$ are imprecise. To overcome this limitation,}

%根据审稿人1的comment16， noise frame data 误解了意思。改一下这段描述。
%\textcolor{red}{In the scenario of our privacy-preserving system, most existing tracking approaches cannot be used for the reason that their necessary inputs are encrypted or encoded.}
%\textcolor{black}{In the scenario of privacy-preserving system, most existing tracking approaches cannot be used for the reason that their necessary inputs are inherently distorted by encryption. Relying on the imprecise bounding boxes obtained from the rough segmentation result,  those traditional tracking-by-detection methods often fail to catch up with the motion and provide a wrong prediction information for clustering. Thus, we propose a refinement strategy and adopt adaptive Kalman filtering for motion tracking. With this novel refinement procedure, we can effectively correct the distorted and noised bounding boxes observed from encrypted videos according to object history motion information.}
Due to the encryption performed on the video bitstream, the video frames of the decompressed video are inherently distorted.
The pixel values in the decompressed encrypted video frame are different from those of the original video.
For example, we cannot observe human beings in the encrypted frames in Fig. 5(b).
The proposed DNRC feature is extracted in the encrypted and compressed domain.
Thus, it may not have enough information to reflect the exact object movements accurately.
The clustering algorithm mainly relies on the isolated frame features, resulting in that the initial motion boxes $\{ {B}_{i,k}^{l} \}_l$ are imprecise.
There are some ``noise'' regions that may be false positively treated as moving objects.
To overcome this limitation, we propose a Kalman-filter based scheme over the feature time sequence to refine the detection results and perform object tracking.
Conventional tracking algorithms on unencrypted videos can easily access the exact pixel values and trust the input bounding boxes.
For example, the related scheme~\citep{WENG20061190}
takes bounding boxes observed in unencrypted video as input and then design a Kalman-filter method to track moving objects.
In their method, they treat the bounding boxes as correct input without any refinements.
When applying this tracking method to the imprecise bounding boxes obtained from encrypted videos, the tracking algorithm is not capable for refining the bounding boxes and cannot have satisfactory tracking results.
Different from the existing track methods, the novelty of our tracking algorithm can be listed as follows.
we propose to adaptively update the measured bounding boxes in the Kalman filtering model to adapt to our application scenario.
We also propose a specified refinement method to update the measured bounding boxes.

As shown in Fig.~\ref{fig:sketch},
the proposed scheme is implemented frame by frame.
In the proposed Kalman filter model, we consider the initial bounding boxes $\{ {B}_{i,k}^{l} \}_l$ as the rough observation data.
Before using them to update the Kalman filter parameters, we modify $\{ {B}_{i,k}^{l} \}_l$ to obtain refined bounding boxes $\{ {B}_{r,k}^{l} \}_l$
according to different cases.
In the $k$-th frame, by synthesizing the feature information in the previous frames, the Kalman filter model outputs the final bounding boxes $\{ {B}_{t,k}^{l} \}_l$, which will be used to improve the initial motion segmentation in the ($k$-1)-th frame.
%A diagram of this tracking method is given in  Fig.~\ref{fig:sketch}.
We sometimes omit the superscript $l$ in the notions ${B}_{i,k}^{l}$, $ {B}_{r,k}^{l} $, and $ {B}_{t,k}^{l} $ for convenience.
Further details of the motion model, object matching metric, and box refinement process are given below.

\subsubsection{Motion Estimation Model}
In Kalman filtering, the \mbox{$k$-th} state vector ${\mathbf{x}}_k$ and the measurement vector ${\mathbf{z}}_k$ can be
computed as
\begin{align}
{\mathbf{x}}_k &={A_k} {\mathbf{x}}_{k-1} + {\mathbf{w}}_k \label{eq:motionEq} \\
{\mathbf{z}}_{k} &= H_k {\mathbf{x}}_{k} + {\mathbf{v}}_{k} \label{eq:measureEq}
\end{align}
where $A_k$ and $H_k$ denote the transition and the measurement matrices, respectively, and ${\mathbf{w}}_k$ and ${\mathbf{v}}_{k}$ are the noise vectors that are drawn from zero-mean multivariate normal distributions with covariances $Q_k$ and $S_k$, respectively.
The actual state of ${\mathbf{x}}_k$ is the internal and hidden variable, which is recursively estimated via the
prediction and the update phases in the Kalman filter.
The prediction phase can be expressed as
\begin{align}
{\mathbf{x}}_{k}^{\mathsmaller -} &= {A_k}{\mathbf{x}}_{k-1}^{\mathsmaller +}  \\
P_{k}^{\mathsmaller -} &= {A_k} P_{k-1}^{\mathsmaller +} A^{\top}_k + Q_k
\end{align}
where ${\mathbf{x}}_{k}^{\mathsmaller -}$ and $P_{k}^{\mathsmaller -}$ are the $k$-th priori estimates
of ${\mathbf{x}}_k$ and the covariance matrix, respectively,
and
${\mathbf{x}}_{k-1}^{\mathsmaller +}$ and $P_{k-1}^{\mathsmaller +}$ are the ($k$-1)-th posteriori estimates of ${\mathbf{x}}_{k-1}$ and the covariance matrix, respectively.
The update phase is formulated as
\begin{align}
K_{k} &= P_{k}^{-}H_{k}^{\top}(H_{k}P_{k}^{-}H_{k}^{\top}+S_{k})^{-1}& \\
{\mathbf{x}}_{k}^{+} &= {\mathbf{x}}_{k}^{-}+K_{k}({\mathbf{z}}_{k}-H_{k} {\mathbf{x}}_{k}^{-}) \label{eq:updateZk}\\
P_{k}^{+} &= (I-K_{k}H_{k})P_{k}^{-}
\end{align}
where $K_k$ denotes the optimal Kalman gain.
For a more detailed explanation of the above notations, please refer to~\citep{welch1995introduction}.

We use $\varphi(\cdot)$ to denote the operator that maps a rectangular box ${{B}}$ to a 4D vector ${\bf z}$ consists of the centroid coordinates $c_{x}$ and $c_{y}$,
the width $w$, and the height $h$, i.e.,
\begin{equation}
\varphi({B}) = {\bf z} = [c_{x}, c_{y}, w, h].
\end{equation}
Conversely, given a 4D vector $\mathbf{z}$, we can uniquely identify a bounding box ${B}$ of which the centroid is determined by the first two elements of $\mathbf{z}$, and the width and height are determined by the last two elements of $\mathbf{z}$, i.e.,
\begin{equation}
\varphi^{-1}({\bf z}) = {B}.
\end{equation}
%We show the relationship between ${B}$, $\mathbf{z}$, and $\varphi$ in Fig.~\ref{fig:varphi}.
%\begin{figure}[!t]
%	\centering	
%	\includegraphics[width=2.8in]{figure/varphi.png}
%	\caption{The relationship between ${B}$, $\mathbf{z}$, and $\varphi$.}
%	\label{fig:varphi}
%\end{figure}

In our Kalman filter model, the $k$-th measurement vector $\mathbf{z}_k$ is the 4D vector of a bounding box ${B}_k$ from observation, i.e.,
\begin{equation}
\mathbf{z}_k=\varphi({B}_k)=\left[c_{xk}, c_{yk}, w_k, h_k\right].
\end{equation}
Thus, the correctness of $\mathbf{z}_k$ depends on the accuracy of ${B}_k$.
In Section.~\ref{sec:boxRef}, we provide box refinement method to improve further the precision of the initial bounding box ${B}_{i,k}$.
Similar with the method in \citep{5512258},
We use an 8D vector ${\mathbf{x}}_k$ as the $k$-th state vector, i.e.,
\begin{equation}
{\mathbf{x}}_k=\left[c_{xk}, c_{yk}, w_k, h_k, v_{xk}, v_{yk}, v_{wk}, v_{hk}\right]
\end{equation}
where $v_{xk}$ and $v_{yk}$ are the horizontal and vertical velocities of the centroid of ${{B}}_{k}$, respectively,
and $v_{wk}$ and $v_{hk}$ are rates of variation of $w_k$ and $h_k$, respectively.

Since the time interval between adjacent frames is very small, we assume that the moving object's velocity is uniform over a frame interval.
Thus, we have
\begin{align}
A_k&=\left[
\begin{array}{cc}
I_{4 \times 4} & \Delta t \cdot I_{4 \times 4} \\
O_{4 \times 4} & I_{4 \times 4}
\end{array}\right] \triangleq A, \\
H_k&=\left[
\begin{array}{cc}
I_{4 \times 4} & O_{4 \times 4}
\end{array}\right] \triangleq H
\end{align}
where $\Delta t$ denotes the time interval between adjacent frames, and
$I_{4 \times 4}$ and $O_{4 \times 4}$ are the $4\times 4$ unit and null matrices, respectively.

\subsubsection{Object Matching}
Since the video is encrypted in our application scenario, the bounding box may be severely  deformed.
The centroid of the bounding box may not reflect the centroid of the moving object.
Therefore we cannot adopt the conventional Euclidean distance as the matching function.
Since the area of overlap between two regions is a more general metric for representing their relationship, we utilize the region overlap rate in our method.
% as the matching function.
We use $\mathcal{R}(\cdot, \cdot)$ to denote the operator computing overlap rate between two boxes.
More specifically, the overlap rate of two bounding boxes ${{B}}$ and ${{B}}'$ is given as
\begin{equation}
\label{E4}
\mathcal{R}\left({{B}},{{B}}'\right) \triangleq
\frac{\left| {{B}} \bigcap {{B}}' \right|} %\Big /
{\left| {{B}} \bigcup {{B}}' \right|}
\end{equation}
where $\bigcap$ and $\bigcup$ are the region union and intersection operations, respectively,
and $|\cdot|$ denotes the element number of the region.

Based on the priori estimate state ${\mathbf{x}}_{k}^{-}$, we can obtain the corresponding priori estimated bounding box ${{B}}_{e,k}^{-}$ determined by $H{\mathbf{x}}_{k}^{-}$, i.e.,
\begin{equation}
{{B}}_{e,k}^{-} = \varphi^{-1} (H{\mathbf{x}}_{k}^{-}).
\end{equation}
For every ${{B}}_{e,k}^{-}$, we can find the ${{B}_{i,k}}$ that has the largest overlap rate among all the initial bounding boxes.
For convenience, we denote $\mathcal{R}({{B}_{i,k}},{{B}}_{e,k}^{-})$ by ${R}_{ie,k}$.
If ${R}_{ie,k} > 0$, we say that this ${{B}_{i,k}}$ \emph{is matched with} ${{B}}_{e,k}^{-}$.
There are two cases when ${R}_{ie,k} = 0$.
\begin{enumerate}
	\item[(i)] When a new initial bounding box is not matched with any existing priori estimated bounding boxes, we create a new tracker for it, and check whether it persists over several frames to avoid tracking noise clusters.
	\item[(ii)] When no initial bounding box is assigned to the existing object, the only possible events for the object are disappearance and stasis.
	Therefore, we remove the tracker of any object that has been lost for too long.
	
\end{enumerate}

\subsubsection{Refinement}\label{sec:boxRef}
%
%\textcolor[rgb]{1.00,0.00,0.00}{Since we perform motion detection and tracking on the encrypted and compressed video bitstream,
%the inherent distortion of encryption and compression will decrease the detection and tracking accuracy.}
To further improve the detection and tracking accuracy,
%solve this problem,
we adaptively refine the measurement in the updating process of the Kalman filter model, based on the overlap rate and the correspondence between ${{B}_{i,k}}$ and ${{B}}_{e,k}^{-}$.
Specifically, for one object with ${{B}}_{e,k}^{-}$, we say that its matching ${{B}_{i,k}}$ is \emph{precise} if ${R}_{ie,k} \geq T_{ie} $, where $T_{ie} $ is a pre-set threshold.
When ${{B}_{i,k}}$ is precise, we will let ${B}_{r,k}={B}_{i,k}$ and convert it to the measurement vector as ${\mathbf{z}}_{k}= \varphi ({B}_{r,k}) = \varphi ({B}_{i,k})$, and set the final detected bounding box
${B}_{t,k}={B}_{i,k}$.

\begin{figure}
	\centering	\includegraphics[width=3.0in]{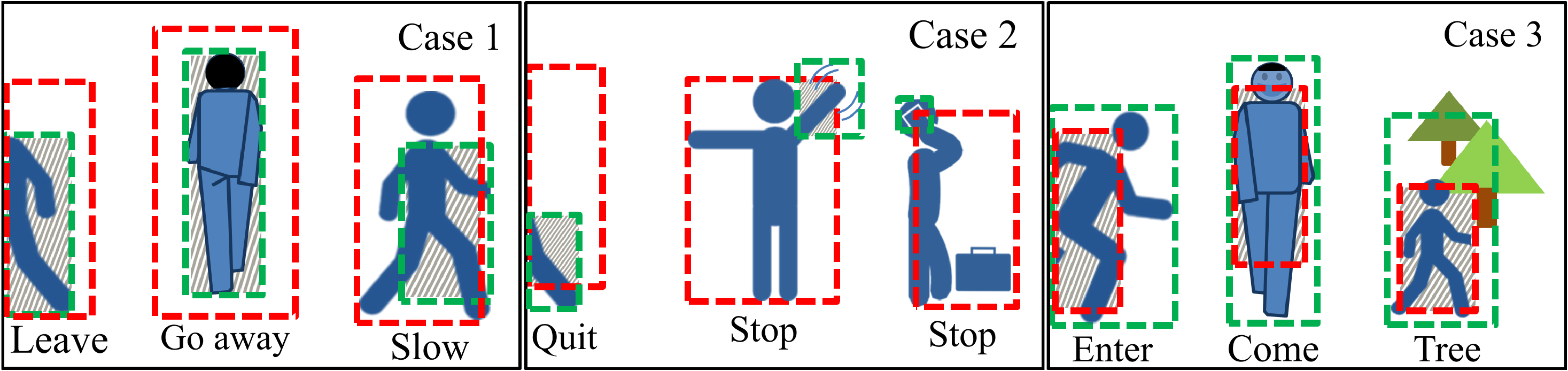}
	\caption{The three cases in which $0 < {R}_{ie,t} < T_{ie}$.
		Green and red dotted lines show the borders of the measured and estimated boxes, respectively. The shadow regions denote the overlapping areas.}
	\label{TrackingExplain}
\end{figure}

\begin{figure}
	\centering	\includegraphics[width=5in]{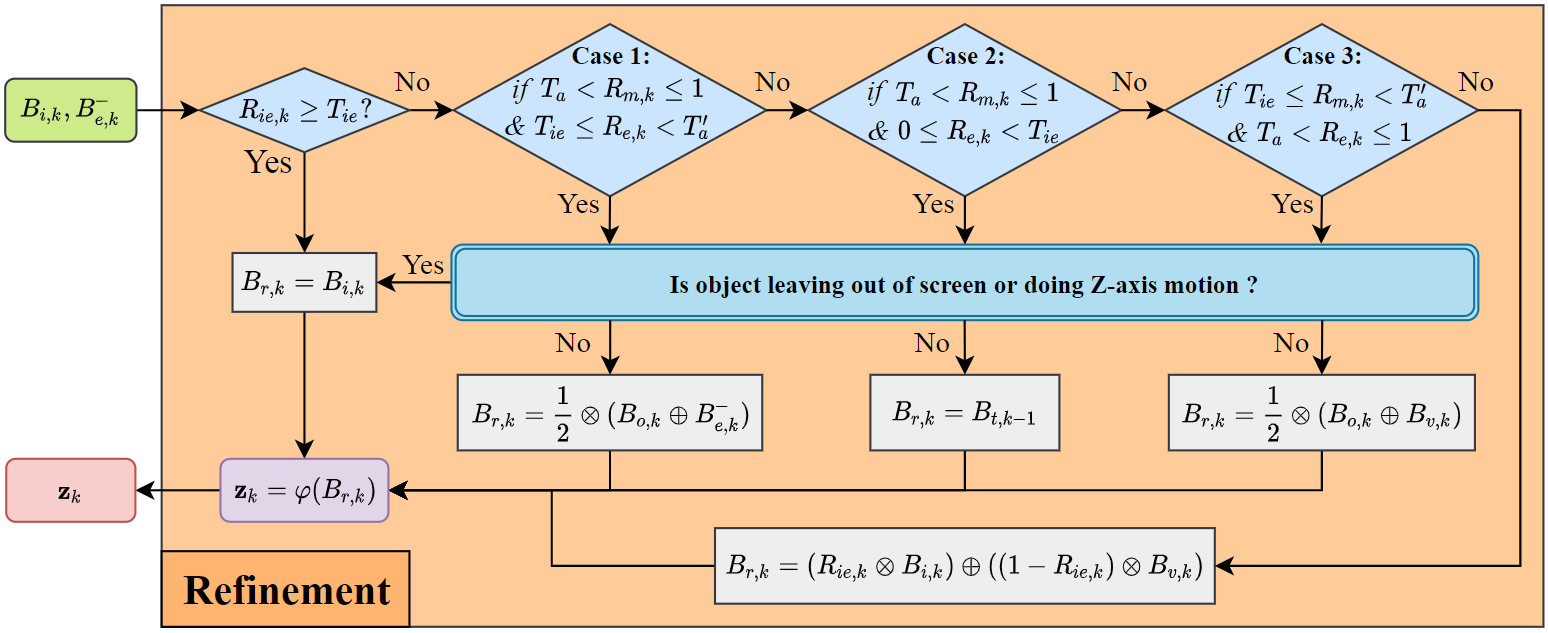}
	\caption{The diagram of the refinement process in our tracking algorithm.}
	\label{refinement_diagram}
\end{figure}

For an object with an imprecise ${{B}_{i,k}}$, i.e., $0 < {R}_{ie,k} < T_{ie}$, we cannot input it to the Kalman function before box refinement.
The small value of ${R}_{ie,k}$ means that there is insufficient overlap between the two boxes.
We then employ two finer metrics
\begin{align}
\label{E5}
{R}_{m,k}&=
%\frac
{\left| {{B}_{i,k}} \bigcap {{B}}_{e,k}^{-} \right|} \Big/ {\left| {{B}_{i,k}} \right|}, \\
{R}_{e,k}&=
%\frac
{\left|  {{B}_{i,k}} \bigcap {{B}}_{e,k}^{-} \right|} \Big/ {| {{B}}_{e,k}^{-} |}
\end{align}
to evaluate the extent to which the overlapping region fits the measured or estimated boxes.
Based on the condition of ${R}_{ie,k}< T_{ie}$, if we limit
one of the metrics to be greater than a threshold $T_{a}$,
then the other will be correspondingly smaller than a threshold $T_{a}^{\prime}$, where
\begin{equation}
T_{a}^{\prime}=
\frac{T_{ie} T_{a}}{T_{ie} T_{a}+T_{a}-T_{ie}.}
\end{equation}
By combining Eq.~(\ref{E4}) and Eq.~(\ref{E5}), we can have that $T_{a}>T_{ie}$ and $T_{a}^{\prime}>T_{ie}$.
For two boxes ${B}$ and ${B}^{\prime}$ with $\varphi ({{B}}) = [c_{x}, c_{y}, w, h] $ and
$\varphi ({B}^{\prime}) = [c'_{x}, c'_{y}, w', h']$,
we define the operations of $\oplus$ and $\otimes$ as
\begin{align}
{{B}} \oplus {{B}}' & = \varphi^{-1} ([ c_{x}+c'_{x},c_{y1}+c'_{y},w+w',h+h' ]), \\
a \otimes {{B}}_{1}  & = \varphi^{-1} ([ac_{x}, ac_{y}, aw, ah]),~~ \forall a \in \mathbb{R}.
\end{align}
Suppose that $ (\overline{v}_{xk}, \overline{v}_{yk})$ is the average velocity of ${B}_{e,k}^{-}$s of one object over the past $\mu$ frames,
$\ (\overline{w}_k, \overline{h}_k)$ is their average size,
and $ (\breve{c}_{xk}, \breve{c}_{yk})$ is the centroid of the bounding box in the latest frame.
The mean size box ${B}_{v,k}$ regarding ${B}_{e,k}^{-}$ is defined as
\begin{equation}
{B}_{v,k} = \varphi^{-1} ([ \breve{c}_{xk}, \breve{c}_{yk}, \overline{v}_{xk}+\overline{w}_k, \overline{vk}_{y}+\overline{h}_k ]).
\end{equation}
We denote the overlap box between ${{B}_{i,k}}$ and ${{B}}_{e,k}^{-} $ by
${{B}}_{o,k} $.
A binary variable $\lambda$ is used to determine the final detection bounding box when ${{B}_{i,k}}$ is not precise.
We illustrate the three possible cases of situation $0 < {R}_{ie,k} < T_{ie}$ in Fig.~\ref{TrackingExplain},
where Case 1 denotes that the area of measurement gets smaller, Case 2 denotes that the area gets too smaller, and Case 3 denotes that the area is growing.
We summarize the refinement process in Fig.~\ref{refinement_diagram} and provide the details in the following.

\begin{itemize}[leftmargin=1em]
	\item \emph{Case 1}: $T_{a}<{R}_{m,k} \leq 1$ and $T_{ie} \leq {R}_{e,k} < T_{a}'$
	
	If the moving object leaves the screen or is undergoing motion along the z-axis, we consider ${{B}_{i,k}}$ to be precise and let ${{B}}_{r,k} = {{B}_{i,k}} $.
	Otherwise, we can adopt the refined bounding box as ${{B}}_{r,k}= \frac{1}{2} \otimes  ({{B}}_{o,k} \oplus {{B}}_{e,k}^{-} )$ and set $\lambda = 0$.

	\item \emph{Case 2}: $T_{a}<{R}_{m,k} \leq 1$ and $0 \leq {R}_{e,k} < T_{ie}$
	
	When the detected object moves out of the screen, we consider ${{B}_{i,k}}$ to be precise and adopt ${{B}}_{r,k} = {{B}_{i,k}} $; otherwise, the object remains in stasis.
	We refine the bounding box as ${{B}}_{r,k} =  {{B}}_{t,k-1}$ and set $\lambda = 0$.

	\item \emph{Case 3}: $T_{ie} \leq {R}_{m,k} < T_{a}^{\prime}$ and $T_{a}<{R}_{e,k} \leq 1$
	
	If the object undergoes motion along the z-axis towards the camera or enters the field of view,
	we consider ${{B}_{i,k}}$ is precise and set ${{B}}_{r,k}= {{B}_{i,k}} $.
	Otherwise, the initial detected bounding box may be merged with background noise, such as noise from trees.
	We employ the refined bounding box as ${{B}}_{r,k} =  \frac{1}{2} \otimes ({{B}}_{o,k} \oplus {{B}}_{v,k} ) $ and set $\lambda = 1$.

	\item \emph{Otherwise}:
	
	If ${R}_{m,k}$ and ${R}_{e,k}$ are too small, the estimation and the observation are too mismatched that we have to discard them both.
	We refine the bounding box as
	\begin{equation}
	{B}_{r,k} = \left({R}_{ie,k} \otimes {{B}_{i,k}} \right) \oplus \left(\left(1-  {R}_{ie,k} \right) \otimes {{B}}_{v,k} \right).
	\end{equation}
	If $\mathcal{R} ({B}_{r,k},{{B}}_{e,k}^{-} ) \geq T_{ie}$,
	we set $\lambda = 1$; otherwise, we set $\lambda = 0$.
\end{itemize}
The refined measurement vector is computed as ${\bf z}_{k} = \varphi({B}_{r,k})$.
At each iteration, ${\mathbf{z}}_{k}$ is passed as feedback to Eq.~(\ref{eq:updateZk}) in the update process, and the subsequent Kalman filtering process is continued.
If ${B}_{i,k}$ is precise, the final detected bounding box ${{B}}_{t,k}$ can be formulated as
\begin{equation}
\label{eq:btk}
{{B}}_{t,k}=
({R}_{ie,k} {{B}_{i,k}} ) \oplus ( (1-{R}_{ie,k} ) {{B}}_{e,k}^+ )
\end{equation}
where ${{B}}_{e,k}^{+}$ denotes the bounding box determined by the posteriori estimate ${\mathbf{x}}_{k}^{+}$, i.e., ${{B}}_{e,k}^{+} = \varphi^{-1} (H{\mathbf{x}}_{k}^{+})$.
Otherwise, the detected bounding box is obtained as
\begin{equation}{\label{eq:btk}}
{B}_{t,k}=\left\{
\begin{array}{l}
{{B}_{e,k}^+,} \qquad {\lambda=0}  \\
{{B}_{r,k},} \qquad {\lambda=1}  \\
\end{array}\right.
\end{equation}

\subsubsection{Occlusion Problem}
The occlusion here means multiple objects are obscured, resulting in their clusters connected in the feature image.
It is difficult to split them in the encrypted and compressed domain.
Because the time interval of two adjacent frames is very short, it is reasonable to assume that
the velocities and the sizes of the bounding boxes of the occluded objects remain unchanged during the occlusion.
In the $k$-th frame, we first compute all the mean size box ${B}_{v,k}$s.
For every object, we delete the points in the feature image $F(k)$ that do not belong to ${B}_{m,k}$ or ${B}_{v,k}$, and then restore the points in the ${B}_{v,k}$ to obtain a new feature image $\tilde F(k)$.
We perform the clustering algorithm on $\tilde F(k)$ and denote the rectangle generated from the biggest cluster as ${B}_{n,k}$.
If $\mathcal{R} ({B}_{n,k},{B}_{e,k}^- ) > T_{ie}$, we let ${\bf{z}}_{k}=\varphi (\frac{1}{2} \otimes ({{B}}_{n,k} \oplus {{B}}_{v,k}))$, otherwise we let ${\bf z}_k = \varphi ({B}_{v,k})$.
The final detected bounding box is ${B}_{t,k}={B}_{e,k}^+$.

\section{Experimental Results}\label{experiment}
In our experiments, we mainly considered surveillance scenarios with low crowd densities, since they are typical of real-world monitoring scenarios with privacy concerns, such as in the home or laboratory.
Our experimental database consisted of 16 video sequences selected from four video data sets: LASIESTA \citep{CUEVAS2016103}, CWD-2014 \citep{6238919}, OTCBVS \citep{davis2007otcbvs}, and CAVIAR \citep{fisher2004caviar}.
These videos contain critical challenges in detection, and we list them in Table \ref{Tabeldatabase}.
Since OTCBVS and CAVIAR contain only bounding box ground truths, we used them in the experiments on motion tracking.
\begin{table}[!t]
	\caption{Video datasets in our experiments.}
	\label{Tabeldatabase}
	\centering
	\renewcommand{\arraystretch}{1.05}
	\begin{tabular}{c|c|c|c|c}
		\hline
		Dataset         & Frames                                            & \begin{tabular}[c]{@{}c@{}} Video\\ number\end{tabular} & Resolution & Challenges                                                         \\
		\hline\hline
		LASIESTA&3931&10&352*288&\begin{tabular}[c]{@{}c@{}}shadows, bootstrap,\\camera motion/jitter,\\dynamic background,\\ temporally static \end{tabular}\\
		\hline
		CDW-2014&800&1&360*240&\begin{tabular}[c]{@{}c@{}}occlusion, shadows \end{tabular}\\
		\hline
		OTCBVS&3068&2&320*240&\begin{tabular}[c]{@{}c@{}}tiny object \end{tabular}\\
		\hline
		CAVIAR&2842&3&384*288&\begin{tabular}[c]{@{}c@{}}temporally static,\\ occlusion, shadows \end{tabular}\\
		\hline
	\end{tabular}
\end{table}

\begin{table}
	\caption{Average results of motion segmentation in the encrypted domain}
	\label{SegmentPerformance}
	\centering
	\renewcommand{\arraystretch}{1.1}
	\begin{tabular}{c|ccc}
		\hline
		Method & Pr & Re & F1\\
		\hline
		LMVD \citep{10.1145/3131342} &0.3823&\textbf{0.9154}&0.4940\\
		CB+PD \citep{8451279} &0.4980&0.8354&0.6148\\
		DNRC (proposed) & \textbf{0.6132}&0.8655&\textbf{0.7083}\\
		\hline
	\end{tabular}
\end{table}

\begin{figure*}[htb]
	\centering
	\includegraphics[width=6.4in]{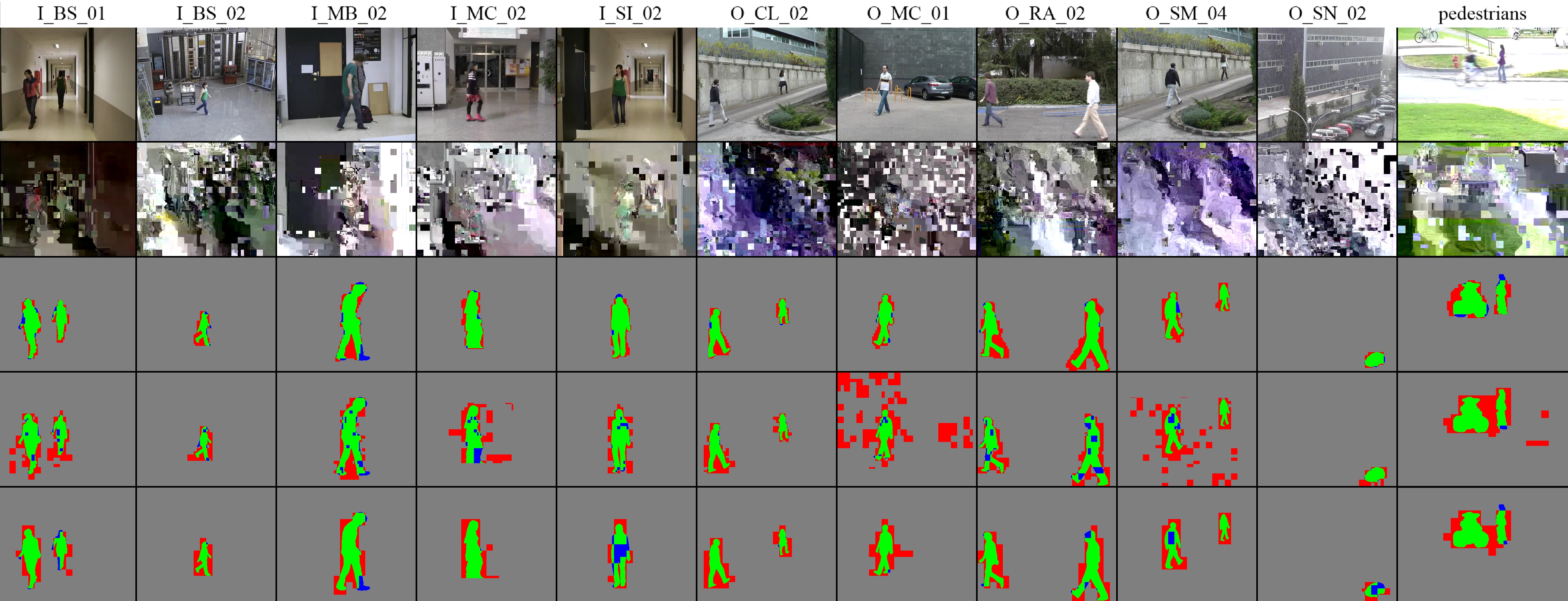}
	\caption{Segmentation results. The top row shows the original frame and the second row is the encrypted frame. The third to the fifth rows show the detection outputs of our proposed algorithm, Guo's algorithm \citep{10.1145/3131342} and Ma's algorithm \citep{8451279}, respectively. Green pixels represent true positive pixels, while blue and red represent false positive and false negative pixels, respectively.}
	\label{segmentationexample}
\end{figure*}

\begin{table*}[!t]
	\caption{Results of motion segmentation in the encrypted  domain}
	\label{SegmentPerformance}
	\centering
	\renewcommand{\arraystretch}{1.1}
	\begin{adjustbox}{max width=0.8\textwidth}
		\begin{tabular}{ccccccccccccc}
			\hline
			\multirow{2}{*}{\begin{tabular}[c]{@{}c@{}}Video name\end{tabular}} && \multicolumn{3}{c}{DNRC (proposed)} && \multicolumn{3}{c}{LMVD \citep{10.1145/3131342}} && \multicolumn{3}{c}{CB+PD \citep{8451279}} \\
			\cline{3-5} \cline{7-9} \cline{11-13}
			&&  Pr  & Re   & F1     &&  Pr  & Re   & F1   &&  Pr  & Re   & F1   \\
			\hline
			I\_BS\_01&&\textbf{0.7059}&0.8650&\textbf{0.7774} &&0.4419&\textbf{0.9517}&0.6036 &&0.5183&0.8804&0.6525 \\
			I\_BS\_02&&\textbf{0.5319}&0.7139&\textbf{0.6096} &&0.4425&\textbf{0.7270}&0.5502 &&0.4601&0.7207&0.5616 \\
			I\_MB\_02&&\textbf{0.7350}&0.6345&0.6810 &&0.5929&\textbf{0.8849}&\textbf{0.7101} &&0.6264&0.7014&0.6617 \\
			I\_MC\_02&&\textbf{0.5326}&0.9428&\textbf{0.6807} &&0.0677&\textbf{0.9890}&0.1267 &&0.4821&0.9587&0.6416 \\
			I\_SI\_02&&\textbf{0.7269}&0.8933&\textbf{0.8016} &&0.5116&\textbf{0.9349}&0.6613 &&0.5437&0.7957&0.6460 \\
			O\_CL\_02&&\textbf{0.6009}&\textbf{0.9605}&\textbf{0.7393} &&0.5063&0.9341&0.6567 &&0.4461&0.9242&0.6017 \\
			O\_MC\_01&&\textbf{0.5686}&0.9669&\textbf{0.7161} &&0.0400&\textbf{0.9972}&0.0769 &&0.4500&0.9602&0.6128 \\
			O\_RA\_02&&0.5452&0.9293&0.6872 &&\textbf{0.5616}&0.9056&\textbf{0.6933} &&0.5229&\textbf{0.9397}&0.6719 \\
			O\_SM\_04&&\textbf{0.5107}&0.9723&\textbf{0.6696} &&0.0455&\textbf{0.9952}&0.0870 &&0.4272&0.9222&0.5839 \\
			O\_SN\_02&&0.7036&0.7329&\textbf{0.7179} &&0.4967&\textbf{0.8261}&0.6204 &&\textbf{0.9222}&0.4582&0.4785 \\
			pedestrians
			&&\textbf{0.5834}&0.9087&\textbf{0.7106} &&0.4986&0.9232&0.6475 &&0.5006&\textbf{0.9283}&0.6505 \\
			\hline
			Average  &&\textbf{0.6132}&0.8655&\textbf{0.7083} &&0.3823&\textbf{0.9154}&0.4940 &&0.4980&0.8354&0.6148 \\
			\hline
		\end{tabular}
	\end{adjustbox}
\end{table*}

\begin{table*}[htb]
	\begin{adjustbox}{max width=1\textwidth}
		\begin{threeparttable}[]
			\caption{Comparison of segmentation results with several popular methods operating in the plain-text domain}
			\label{Segment_compare_table}
			\renewcommand{\arraystretch}{1.01}
			
			\begin{tabular}{c|c|ccc|ccc|ccc|ccc}
				\hline
				\multirow{3}{*}{\begin{tabular}[c]{@{}c@{}}Algorithm\\ Name\end{tabular}} & \multirow{3}{*}{\begin{tabular}[c]{@{}c@{}}Detector\\ Domain\end{tabular}} & \multicolumn{6}{c|}{Stationary Camera}                                                                         & \multicolumn{6}{c}{Moving Camera}                                                                        \\ \cline{3-14}
				&                                                                            & \multicolumn{3}{c|}{Normal Videos}                  & \multicolumn{3}{c|}{Encrypted Videos}                & \multicolumn{3}{c|}{Normal Videos}                  & \multicolumn{3}{c}{Encrypted Videos}               \\ \cline{3-14}
				&                                                                            & Pr              & Re              & F1              & Pr              & Re               & F1              & Pr              & Re              & F1              & Pr              & Re              & F1              \\ \hline
				GMM\citep{784637}                                                                     & pixel                                                                      & 0.6368          & 0.8393          & 0.7104          & 0.0289          & 0.5360           & 0.0508          & 0.3325          & 0.7316          & 0.3754          & 0.0217          & 0.5133          & 0.0415          \\
				SC-SOBS\citep{6238922}                                                                   & pixel                                                                      & 0.7746          & 0.7370          & 0.7329          & 0.0300          & 0.8787           & 0.05745         & 0.2455          & 0.8528          & 0.3077          & 0.0227          & 0.9082          & 0.0441          \\
				SuBSENCE\citep{6975239}                                                                  & pixel                                                                      & \textbf{0.9393} & 0.7056          & \textbf{0.7925} & 0.0425          & 0.1763           & 0.0661          & \textbf{0.6316} & 0.84597         & 0.6263          & 0.0472          & 0.1917          & 0.0737          \\
				ST-MRF\citep{6272352}                                                                    & CVB\tnote{1}                                                                       & 0.7000          & 0.5029          & 0.5584          & \textbf{0.6693} & 0.1089           & 0.1820          & 0.5547          & 0.72277         & 0.6134          & 0.1720          & 0.0804          & 0.0663          \\
				Bhaskar\citep{6490026}                                                                   & CVB                                                                        & 0.6644          & 0.2998          & 0.3991          & 0.0377          & 0.3683           & 0.0653          & 0.5978          & 0.3878          & 0.3909          & 0.0432          & 0.3636          & 0.0722          \\
				Zhao\citep{7801078}                                                                      & CVB                                                                        & 0.5283          & 0.8224          & 0.6394          & 0.1842          & 0.7022           & 0.2816          & 0.0685          & 0.9595          & 0.1244          & 0.0325          & 0.95393         & 0.0624          \\
				LMVD\citep{10.1145/3131342}                                                                       & CVB                                                                        & 0.5065          & \textbf{0.8859} & 0.6429          & 0.50651         & \textbf{0.88594} & 0.6429          & 0.0511          & \textbf{0.9938} & 0.0969          & 0.0511          & \textbf{0.9938} & 0.0969          \\
				CB+PD\citep{8451279}                                                                         & CVB                                                                       & 0.5149          & 0.7936          & 0.6155          & 0.51485         & 0.79356          & 0.6155          & 0.4531          & 0.94703         & 0.6127          & 0.4531          & 0.94703         & 0.6127          \\
				DNRC(proposed)                                                                  & CVB                                                                        & 0.6434          & 0.8298          & 0.7167          & 0.64339         & 0.82976          & \textbf{0.7167} & 0.5405          & 0.9606          & \textbf{0.6915} & \textbf{0.5405} & 0.9606          & \textbf{0.6915} \\ \hline
			\end{tabular}
			
			\begin{tablenotes}
				\item[1] Compressed Video Bitstream
			\end{tablenotes}
			
		\end{threeparttable}
	\end{adjustbox}
\end{table*}

We implemented the proposed encryption scheme using the H.264/AVC reference software JM-19.0.
We used the \emph{high profile} with the GOP structure of IPPP.
All the video sequences were encoded at 30 frames per second with an intra-period of 12 and a fixed quantization step parameter of 28. All the experiments were carried out on a 64-bit Windows 7 PC with a 3.40 GHz Intel Core i7-6700 CPU and 32 GB RAM.
As discussed in~\citep{milan2016mot16}, the overlap rate of a match is at least 0.5. Thus, we set $T_{ie}=0.6$ and $T_{a}=0.8$ to obtain a more credible tracking result, and it can be inferred that $T_{a}^{\prime}\approx 0.7$.
We also set $\mu = 5 $ frames
%and $\alpha = 8 $ blocks %这个参数是用于形态学去噪的，已经去掉了。
for the experiments.

\subsection{Performance of Motion Segmentation}
%In general, the performance of motion segmentation is evaluated by precision (Pr), recall (Re), and F1-score (F1),\citep{6238919} i.e.,
%\begin{gather}
%\mathrm{Pr}=\frac{TP}{TP+FP}\\
%\mathrm{Re}=\frac{TP}{TP+FN}\\
%\mathrm{F1}=\frac{2 \times \mathrm{Re} \times \mathrm{Pr}}{\mathrm{Re} + \mathrm{Pr}}
%\end{gather}
%where $TP$, $FP$ and $FN$ are the pixel number of true positive, false positive, and false negative, respectively.

\begin{figure*}[!t]
	\centering
	\includegraphics[width=5.8in]{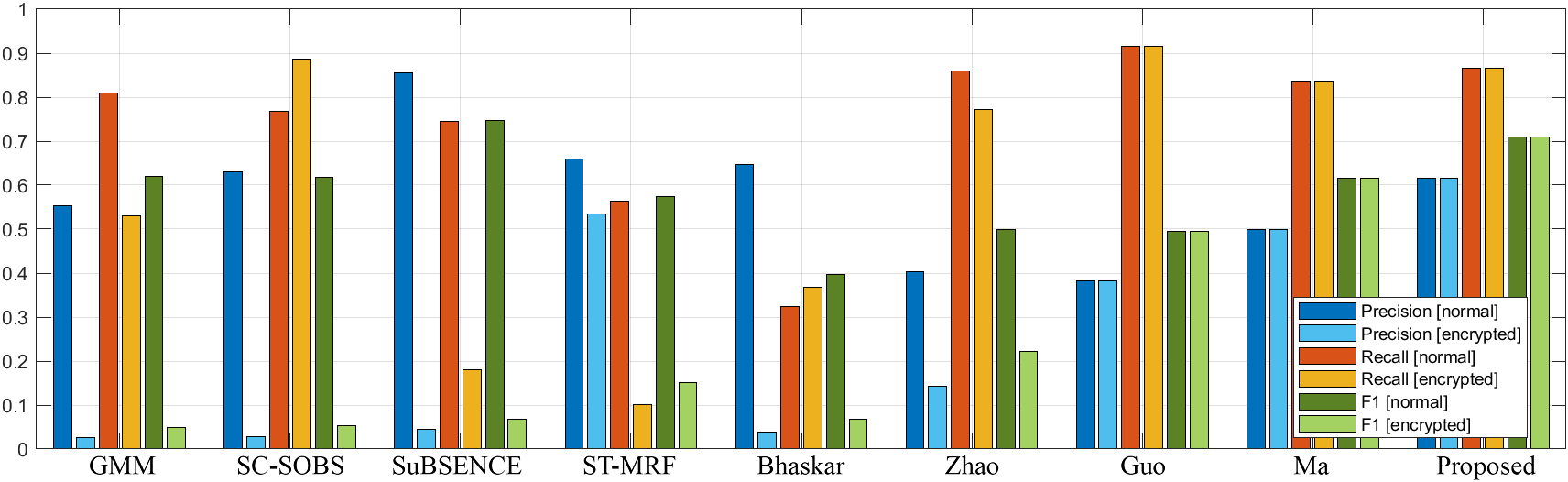}
	\caption{Comparison of average segmentation performance between normal and encrypted videos.}
	\label{Segment_compare_bar}
\end{figure*}

In general, the performance of motion segmentation is evaluated based on the metrics of precision (Pr), recall (Re), and F1-score (F1) \citep{6238919}.
%, i.e.,
%\begin{gather}
%\mathrm{Pr}=\frac{TP}{TP+FP}\\
%\mathrm{Re}=\frac{TP}{TP+FN}\\
%\mathrm{F1}=\frac{2 \times \mathrm{Re} \times \mathrm{Pr}}{\mathrm{Re} + \mathrm{Pr}}
%\end{gather}
%where $TP$, $FP$ and $FN$ are the element number of true positive, false positive, and false negative, respectively.
We compared the segmentation performance of our scheme with other existing schemes \citep{10.1145/3131342,8451279} on encrypted video.
The scheme proposed by Ma \emph{et al}. \citep{8451279} was modified to use H.264/AVC.
The experimental results are shown in Table~\ref{SegmentPerformance},
where it can be seen that our proposed scheme outperformed the other existing schemes on encrypted and compressed video.
Our scheme achieved superior results for precision and F1-score.
Although Guo's scheme \citep{10.1145/3131342} had a higher score for average recall, it contained too many false detections in complex scenarios.
The main reason for this is that moving cameras and dynamic backgrounds can make it challenging to distinguish noise motion vectors from the foregrounds.
However, using the proposed feature, we can easily reduce the background noise.
Because the motion of the camera and the presence of a dynamic background usually cause small and scattered DNRCs.
All of the average values of the segmentation metrics with DNRC were better than those with
CB and PD (CB+PD).
Since the finest scale of our feature is 4$\times$4 block level, compared to 16$\times$16 block level for Ma's and Guo's features, our scheme gives more precise segmentation results.
A comparison of several examples of segmentation results is given in Fig.~\ref{segmentationexample}.
Guo's method shows apparent regions of noise in the \emph{O\_MC\_01} and \emph{O\_SM\_04} sequences.
In Ma's method, the edges of the object regions are rough, and there are many false detections around the object.
In contrast, we can see that our algorithm can accurately detect moving objects without including an additional noise cluster region.
Most regions are classified correctly, and only a few red and blue pixels are seen at the edges of the objects.

\begin{table}
	\caption{Motion detection and tracking results in encrypted domain.}
	\label{TrackingPerformance}
	\centering
	\renewcommand{\arraystretch}{1.1}
	\begin{tabular}{c|cccc}
		\hline
		Method & AUC & Pre20 & MOTA & MOTP\\
		\hline
		LMVD \citep{10.1145/3131342} &0.3688&0.6658&-0.1601&0.6324\\
		CB+PD \citep{8451279} &0.4307&0.7972&-0.0210&0.6613\\
		DNRC (proposed) & \textbf{0.6633}&\textbf{0.9544}&\textbf{0.7230}&\textbf{0.7100}\\
		\hline
	\end{tabular}
\end{table}
\begin{figure*}[!t]
	\centering
	\subfigure{
		\centering
		\includegraphics[width=6.4in]{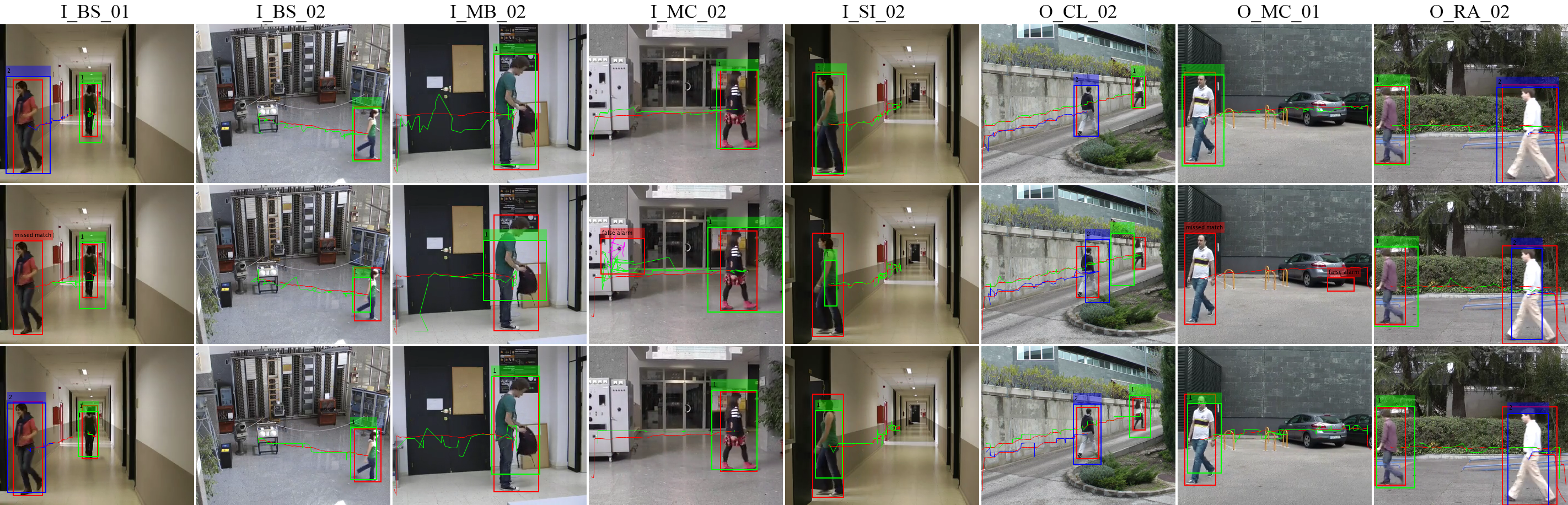}
	}
	\subfigure{
		\centering
		\includegraphics[width=6.4in]{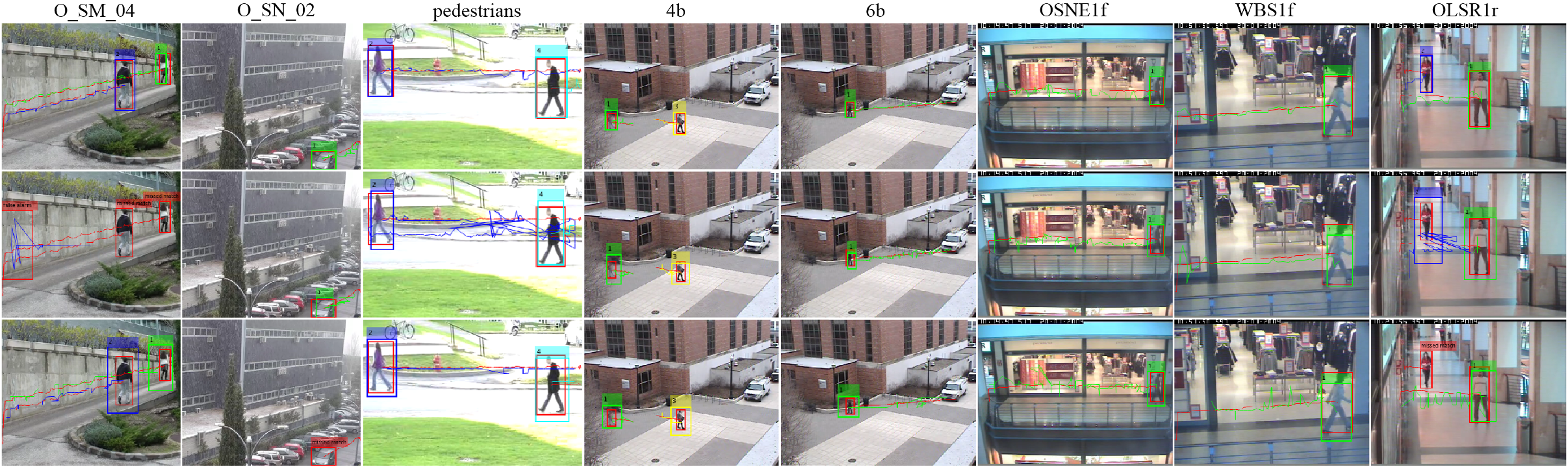}
	}
	\caption{Examples of motion detection and tracking. The first row shows the results from our scheme, while the second and third rows show Guo's \citep{10.1145/3131342} and Ma's \citep{8451279} algorithms, respectively. Bounding boxes and trajectories in red represent the ground truth, while the other colors represent the outputs of the algorithms.}
	\label{CompareInScenario}
	
\end{figure*}

\begin{table*}[htb]
	\caption{Motion detection and tracking results in encrypted domain.}
	\label{TrackingPerformance}
	\centering
	\renewcommand{\arraystretch}{1.0}
	\begin{adjustbox}{max width=1\textwidth}
		\begin{tabular}{lcccccccccccccc}
			\hline
			\multirow{2}{*}{\begin{tabular}[c]{@{}c@{}}Video name\end{tabular}} & \multicolumn{4}{c}{Proposed} && \multicolumn{4}{c}{Guo \citep{10.1145/3131342}} && \multicolumn{4}{c}{Ma \citep{8451279}} \\
			\cline{2-5} \cline{7-10} \cline{12-15}
			& AUC  & Pre20& MOTA & MOTP && AUC  & Pre20& MOTA  & MOTP && AUC  & Pre20&  MOTA & MOTP\\
			\hline
			I\_BS\_01&\textbf{0.7196}&0.9149&\textbf{0.7957}&\textbf{0.7622}&&0.1344&0.3510&-0.6271&0.5686&&0.6127&\textbf{0.9338}& 0.6027&0.6698\\
			I\_BS\_02&0.6428&0.8657&0.5865&\textbf{0.7264}&&0.4843&0.9921&-0.0079&0.5831&&\textbf{0.6722}&\textbf{1.0000}& \textbf{0.7377}&0.7060\\
			I\_MB\_02&0.6002&0.6161&0.4554&0.6903&&0.3588&0.6463&-0.7235&0.5681&&\textbf{0.6722}&\textbf{0.7243}& \textbf{0.8131}&\textbf{0.7030}\\
			I\_MC\_02&\textbf{0.6617}&0.7938&\textbf{0.6495}&\textbf{0.7192}&&0.0722&0.0741&-3.7800&0.0000&&0.6157&\textbf{0.9575}& 0.6170&0.7032\\
			I\_SI\_02&\textbf{0.7781}&\textbf{0.9755}&\textbf{0.9406}&\textbf{0.7995}&&0.4034&0.9340&-0.3251&0.6715&&0.4987&0.9754& 0.0148&0.6380\\
			O\_CL\_02&\textbf{0.7487}&0.9977&\textbf{0.9400}&\textbf{0.7589}&&0.3901&0.7082&-0.0871&0.6861&&0.5559&\textbf{1.0000}& 0.2554&0.6553\\
			O\_MC\_01&\textbf{0.6821}&0.9744&\textbf{0.8564}&\textbf{0.7101}&&0.0112&0.0088&-2.8214&0.6291&&0.5249&\textbf{0.9893}& 0.2366&0.6232\\
			O\_RA\_02&\textbf{0.7082}&0.9097&\textbf{0.7208}&\textbf{0.7639}&&0.6281&0.9595& 0.7432&0.6821&&0.6530&\textbf{0.9608}& 0.7124&0.7114\\
			O\_SM\_04&\textbf{0.5996}&0.9661&\textbf{0.4621}&\textbf{0.6973}&&0.0328&0.0166&-1.8031&0.5688&&0.5564&\textbf{0.9933}& 0.2953&0.6499\\
			O\_SN\_02&\textbf{0.6398}&\textbf{1.0000}&\textbf{0.6356}&\textbf{0.6791}&&0.4761&0.9739&-0.0970&0.5834&&0.0000&0.0000& 0.0000&0.0000\\
			pedestrians
			&\textbf{0.6916}&0.9603&\textbf{0.7339}&\textbf{0.7401}&&0.3556&0.5740& 0.1754&0.6514&&0.6215&\textbf{0.9866}& 0.5898&0.6786\\
			4b       &\textbf{0.5917}&\textbf{1.0000}&\textbf{0.5575}&\textbf{0.6368}&&0.5066&0.9953& 0.0651&0.6249&&0.3110&1.0000&-0.9932&0.5293\\
			6b       &\textbf{0.6642}&0.9882&\textbf{0.8660}&\textbf{0.6898}&&0.5548&\textbf{0.9885}& 0.4163&0.6258&&0.2926&0.7708&-0.5701&0.5334\\
			OSNE1f
			&\textbf{0.6025}&\textbf{1.0000}&\textbf{0.5467}&\textbf{0.6641}&&0.5303&0.9719& 0.2303&0.6082&&0.4590&0.8554&-0.05421&0.6162\\
			WBS1f
			&\textbf{0.7128}&\textbf{0.9614}&\textbf{0.8491}&\textbf{0.7435}&&0.6078&0.8540& 0.5752&0.6519&&0.3030&0.3580&-0.2585&0.6888\\
			OLSR1c
			&\textbf{0.6612}&\textbf{0.9059}&\textbf{0.6849}&\textbf{0.7163}&&0.3192&0.6618&-0.7344&0.5793&&0.3859&0.4927& 0.0027&0.6481\\
			\hline
			Overall  &\textbf{0.6633}&\textbf{0.9544}&\textbf{0.7230}&\textbf{0.7100}&&0.3688&0.6658&-0.1601&0.6324&&0.4307&0.7972&-0.0210&0.6613\\
			\hline
		\end{tabular}
	\end{adjustbox}
\end{table*}
We also compare our scheme with some previous methods \citep{7801078,10.1145/3131342,8451279,784637,6238922,6975239,
	6272352,6490026} on encrypted and unencrypted videos.
As some methods are only suitable for stationary backgrounds, the videos in our dataset were divided into two categories, with stationary and moving cameras.
Since the encrypted domain feature remains unchanged after encryption, algorithms based on this can obtain the same segmentation results on encrypted videos as on the original videos.
We evaluated the segmentation results pixel by pixel rather than block by block, and the experimental results are given in Table~\ref{Segment_compare_table}.
%\st{To our best knowledge, it is impossible to separate a moving object from the background at the pixel scale with a compressed-domain based method.}
%重复语句， reviewer1 的 comment 19
To our best knowledge, it is impossible to separate a moving object from the background at the pixel scale with a compressed-domain based method.
That is because all of the available features extracted from bitstreams have a minimum scale of 4$\times$4 pixels.
Therefore, pixel-domain-based methods generally outperform compressed-domain-based methods on unencrypted video,
except that our method achieved the best F1 performance for unencrypted videos with moving cameras.
However, %as shown in Fig.~\ref{Segment_compare_bar},
the performance of pixel-domain-based methods is sharply reduced when dealing with encrypted video.
Although Guo's scheme obtained the best recall performance for encrypted video with stationary cameras, it ultimately failed to detect objects in videos with moving cameras ($\mbox{Pr} = 0.0511$).
Generally, our scheme has the best F1 performance on encrypted videos among all the compared methods,
whenever the camera is stationary or moving.

\begin{figure}[]
	\centering
	\subfigure[Plaintext]{
		\begin{minipage}[b]{1in}
			\includegraphics[width=1in]{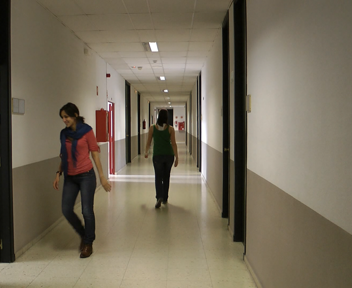}\vspace{4pt}
			\includegraphics[width=1in]{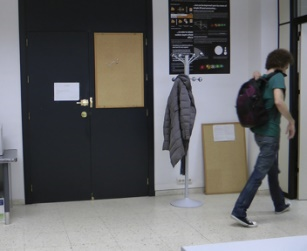}
		\end{minipage}
	}
	\subfigure[Groundtruth]{
		\begin{minipage}[b]{1in}
			\includegraphics[width=1in]{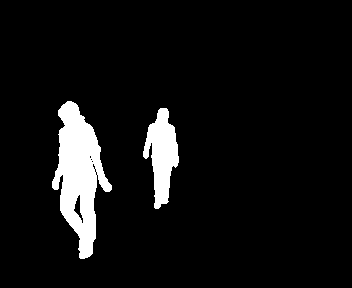}\vspace{4pt}
			\includegraphics[width=1in]{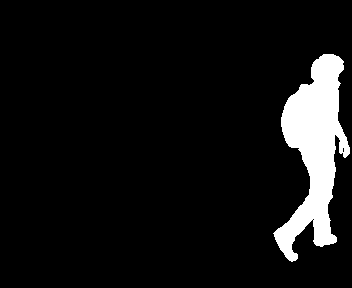}
		\end{minipage}
	}
	\subfigure[GMM]{
		\begin{minipage}[b]{1in}
			\includegraphics[width=1in]{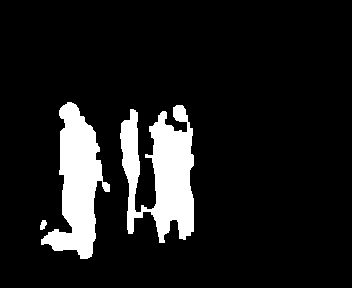}\vspace{4pt}
			\includegraphics[width=1in]{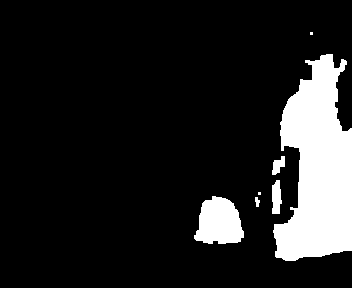}
		\end{minipage}
	}
	\subfigure[Proposed]{
		\begin{minipage}[b]{1in}
			\includegraphics[width=1in]{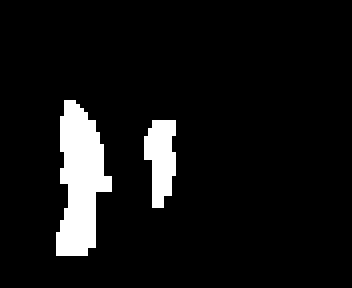}\vspace{4pt}
			\includegraphics[width=1in]{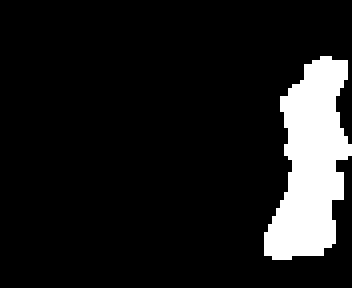}
		\end{minipage}
	}
	\caption{Comparison of motion segmentation. Rows 1 and 2 are the frame \#68 of \emph{I\_BS\_01} video and the frame \#283 of  \emph{I\_MB\_02} video, respectively.}
	\label{GMM}
\end{figure}

%reviewer2 的comment 3
Our performance can catch up with that of some pixel-based methods owing to the proposed DNRC feature and the proposed
refinement method.
Take GMM \citep{784637} as an example of pixel-based methods.
GMM needs to model the background and then find motion regions based on local pixel changes.
Since any pixel change will be reflected in residual data, DNRC can also well reflect where the motion regions are.
Besides, some video sequences do not leave sufficient initial frames for background modeling of GMM when the objects are moving from the first frame.
It results in that GMM detects a ``ghost'' and thus performs poorly in these video sequences.
Two examples of ``ghost'' are shown in Fig. \ref{GMM}.
We can also see that GMM leaves a ``ghost'' when the background object of the backpack is removed.
As for the proposed refinement method, it is used to adaptively update the measured bounding boxes before feeding them into the Kalman-filtering based motion model.
It helps the detector to focus on the candidate motion regions to achieve better clustering results and then improve the detection performance.

\begin{figure}[]
	\centering	
	\subfigure[Success Plots]{
		\centering
		\includegraphics[width=2.3in]{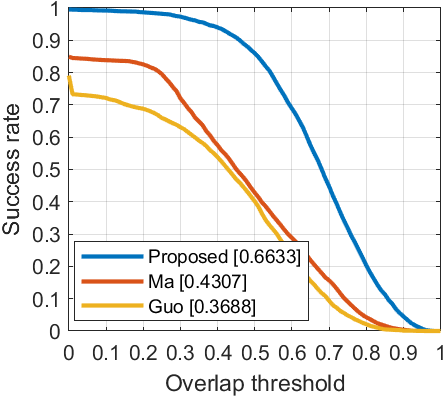}
	}
	\subfigure[Precision Plots]{
		\centering
		\includegraphics[width=2.3in]{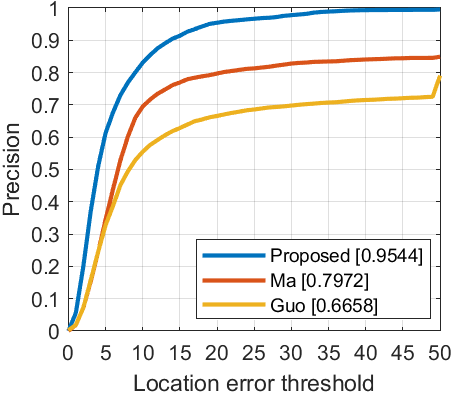}
	}
	\caption{Comparison of success and precision plots between our scheme and the related schemes \citep{10.1145/3131342,8451279} on encrypted videos.}
	\label{SPandPPofMethod}
\end{figure}

\subsection{Performance of Motion Detection and Tracking}
Many previous studies of 2D view-based online object detection and tracking schemes have used success and precision plots to evaluate these schemes against an object tracking benchmark \citep{Wu_2013_CVPR}.
A success plot measures the bounding box overlap, while a precision plot measures the central location error.
Besides these two measures,
we also evaluate the detection and tracking performance using two famous metrics in \citep{milan2016mot16}: multiple object tracking accuracy (MOTA) and multiple object tracking precision (MOTP).
In MOTA and MOTP, the overlap rate of a match is at least 0.5.
%Given the misses box number ($fn_t$), the false detection box number ($fp_t$), switching times of ID ($IDsw_t$), the number of ground truth objects ($G_t$), the bounding box overlap of target $i$ with its assigned ground truth object ($d_{t,i}$), and the number of matches ($c_t$) in frame $t$, we have
%\begin{align}\
%\mathrm{MOTA}&=1-\frac{\sum_{t}\left({fn}_{t}+{fp}_{t}+{IDsw}_{t}\right)}{\sum_t G_t}, \\
%\mathrm{MOTP}&=\frac{\sum_{t,i} d_{t,i}}{\sum_t c_t}.
%\end{align}

We ran a one-pass evaluation (OPE) \citep{Wu_2013_CVPR} from the first frame to the end, since no initialization was used in our moving object tracking task.
Because motion detection and tracking methods based on the pixel domain do not work in our privacy-preserving application scenario, we only carried out a comparison between schemes \citep{10.1145/3131342,8451279} designed for the encrypted and compressed domain.
Some videos in the databases did not contain ground truth bounding boxes (GT-boxes).
We then generated the GT-boxes by drawing the minimum rectangular boxes necessary to bound the objects, strictly respecting the segmentation ground truth in each frame.

The experimental results of motion detection and tracking on encrypted videos are given in Table~\ref{TrackingPerformance}.
Our scheme showed comparable performance for all videos and achieved the best average values for each metric than the other existing methods.
We also plotted the tracking boxes and trajectories in the unencrypted frames to visualize the tracking results, and some examples are shown in Fig. \ref{CompareInScenario}.
Guo's method performed poorly for all videos with a moving camera, due to incorrect detection of background noise.
The inferior trajectory results indicate that this method does not work in complex scenarios.
Ma's scheme also failed to track objects in some videos such as \emph{O\_SN\_02} and \emph{OLSR1r}.
In Ma's method, small-sized objects are hard to detect with 16$\times$16 block-level features, and short trajectories are discarded.
Additionally, the trajectories generated by Ma's scheme contain more fluctuations, meaning that this scheme is unstable.
In contrast, our method can effectively detect small-sized objects and have robust tracking performance on all encrypted videos.
\begin{figure*}[!t]
	\centering
	
	%	\subfigure[from frame 902 to 932 in \emph{4b}]{
	%		\centering
	%		\label{normal_occlusion}
	%		\includegraphics[width=6.0in]{figure/4b_MotionDetectResult_v20190902_all.png}
	%	}
	\subfigure[Frames 416 to 446 from \emph{pedestrians}]{
		\centering
		\includegraphics[width=5.4in]{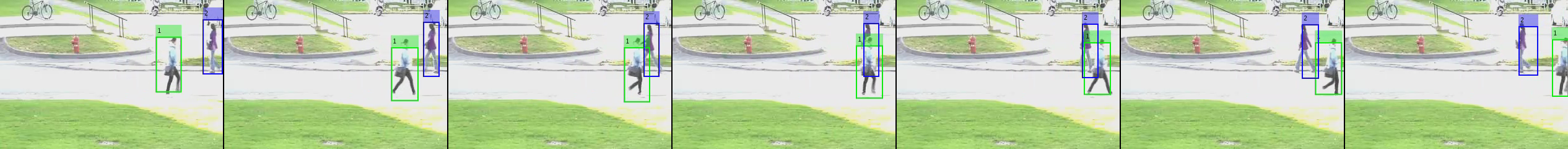}
	}
	\subfigure[Frame 213 to 243 from \emph{OLSR1c}]{
		\centering
		\includegraphics[width=5.4in]{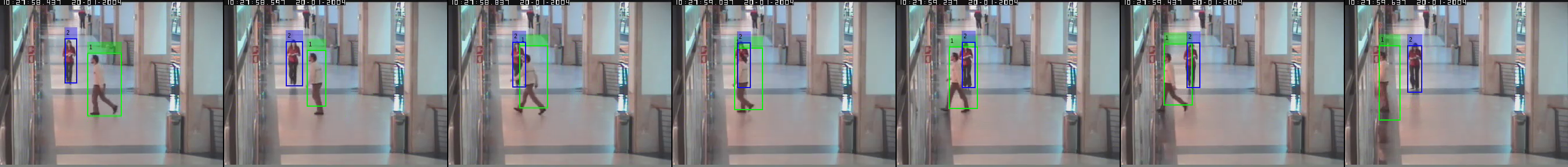}
	}
	\caption{Series of pictures depicting detection and tracking results over a period in which an occlusion occurs. The interval between pictures is five  frames.}
	\label{SpecialOcclusion}
\end{figure*}

\begin{table*}

	\caption{Motion segmentation and tracking results on the encrypted video bitstreams with different encryption schemes.}
	\label{adapttacking}
	\centering
	\renewcommand{\arraystretch}{1.1}
	\begin{adjustbox}{max width=1\textwidth}
		\begin{tabular}{c|c|ccc|cccc}
			\hline
			\multirow{2}{*}{\begin{tabular}[c]{@{}c@{}}Encryption\\ Scheme\end{tabular}} & \multirow{2}{*}{\begin{tabular}[c]{@{}c@{}}Codec\\ Format\end{tabular}} & \multicolumn{3}{c|}{Motion Segmentation}                                                   & \multicolumn{4}{c}{Motion tracking} \\ \cline{3-9}
			&                                                                         & \multicolumn{1}{l}{Precision} & \multicolumn{1}{l}{Recall} & \multicolumn{1}{l|}{F1-score} & AUC     & PRE20   & MOTA   & MOTP   \\ \hline
			Xu \citep{6725633}                                                                    & H.264/AVC                                                               & 0.6132                        & 0.8655                     & 0.7083                        & 0.6633  & 0.9544  & 0.7230 & 0.7100 \\
			Shahid \citep{5733402}                                                                & H.264/AVC                                                               & 0.6132                        & 0.8655                     & 0.7083                        & 0.6633  & 0.9544  & 0.7230 & 0.7100 \\
			Sallam \citep{8119905}                                                                & H.265/HEVC                                                              & 0.6435                        & 0.8374                     & 0.7178                        & 0.6733  & 0.8908  & 0.6930 & 0.7402 \\
			Shahid \citep{6589970}                                                                & H.265/HEVC                                                              & 0.6435                        & 0.8374                     & 0.7178                        & 0.6733  & 0.8908  & 0.6930 & 0.7402 \\ \hline
		\end{tabular}
	\end{adjustbox}
\end{table*}

We also illustrate our scheme's tracking performance in terms of handling an occlusion challenge using two examples, as shown in Fig. \ref{SpecialOcclusion}.
Although most regions of the occluded object are obscured by another moving object, the proposed algorithm can assign their IDs correctly and track them properly.
Thus, our scheme allows for reliable tracking of the detected objects even under occlusion.

To further demonstrate the detection and tracking performance of our scheme, Fig. \ref{SPandPPofMethod} shows success and precision plots for all videos in comparison with the other methods.
These results show that the detection and tracking performance of the proposed algorithm is significantly better than the other tracking methods.

%To investigate the improvement of our tracking algorithm, we also conduct an experiment for the comparison between the conventional tracking algorithm based on Kalman filter \citep{WENG20061190} and the proposed tracker. The results in TABLE R1-10 demonstrate that the proposed algorithm can significantly improve the tracking performance.
%\begin{table}[h]
%	\caption{Comparison of tracking performance on different methods.}
%	\label{HDtable}
%	\centering
%	\renewcommand{\arraystretch}{1.05}
%	\begin{tabular}{l|llll}
%		\hline
%		Method           & AUC    & PRE20  & MOTA    & MOTP   \\ \hline
%		proposed         & 0.6633 & 0.9544 & 0.7230  & 0.7100 \\
%		Kalman \citep{welch1995introduction} & 0.2825 & 0.4044 & -1.5341 & 0.7067 \\ \hline
%	\end{tabular}
%\end{table}

\subsection{Performance on High-Definition Video}
The proposed feature has the advantage of detecting more tiny objects from HD surveillance video and therefore obtains a good performance.
We conducted HD video experiments to compare the performance between our 4$\times$4 feature and 16$\times$16 features.
To get a residual feature with a scale of 16$\times$16 pixels, we extract the number of non-zero residual coefficients in a macroblock, denoted as DNRC\_MB.
We also compare with LMVD~\citep{10.1145/3131342} and CB+PD~\citep{8451279}.
Some snippets with 1920$\times$1080 resolution from VIRAT \citep{5995586} are used for testing.
The experimental results are summarized in Table \ref{HDtable}.
We can see that our 4$\times$4 feature performs better than all the compared 16$\times$16 features for HD videos.

\begin{table}[!t]
	\caption{Tracking performance on HD video using different scale features.}
	\label{HDtable}
	\centering
	\renewcommand{\arraystretch}{1.05}
\begin{tabular}{c|cccc}
	\hline
	Feature Name    & AUC             & PRE20           & MOTA            & MOTP            \\ \hline
	DNRC (proposed) & \textbf{0.6202} & \textbf{0.9802} & \textbf{0.4688} & \textbf{0.6883} \\
	DNRC\_MB        & 0.5878          & 0.9520          & 0.3375          & 0.6691          \\
	LMVD~\citep{10.1145/3131342}  & 0.4695          & 0.7799          & 0.3856          & 0.6472          \\
	CB+PD~\citep{8451279} & 0.5636          & 0.9318          & 0.4515          & 0.6360          \\ \hline
\end{tabular}
\end{table}

\section{Discussions}\label{discussion}

\subsection{Explanation of the Proposed Feature}
In video encoding frameworks, prediction and compensation technologies are widely adopted to reduce spatial and temporal redundancies.
In general, the pixel values in the background area are similar, since most of the background is smooth, without a complicated context.
Therefore, the prediction of the background will result in small residues of blocks in the background regions.
However, in a video with a moving camera, the compensation for this motion can significantly reduce the amount of residual data.
In contrast, the moving object region has a more sophisticated level of detail.
The pixel values in the moving object region change rapidly, reflecting the object's motion, which cannot be predicted by the video encoder.
Thus, the prediction and compensation of the moving object regions will require a more significant amount of residual data to represent the details and the unexpected motion.
As a measure of the amount of residual data, the proposed DNRC feature can reflect the moving object region.

\begin{figure}[!t]
	\centering
	\includegraphics[width=3.5in]{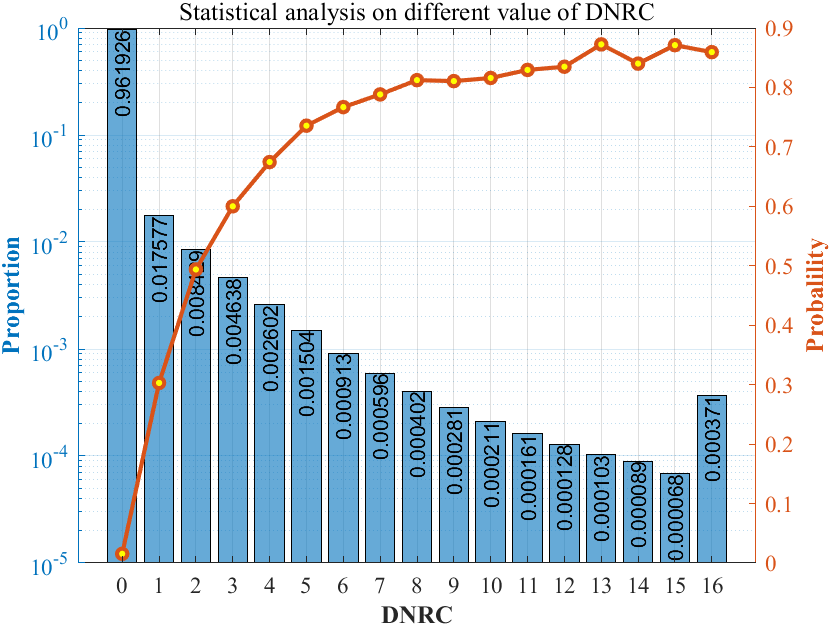}
	\caption{Statistical analysis of values of DNRC: the blue bars indicate the distribution of DNRC in more than 10,000 frames, and the proportion (percentage) of each DNRC value is shown in the corresponding bar. The orange curve shows the statistical probability that each block of DNRC values is a motion block.}
	\label{DNRC_Analysis}
\end{figure}

We used 10,324 labeled frames (except I frame) to analyze the raw DNRC data's statistical characteristics and show the results in Fig.~\ref{DNRC_Analysis}.
The curve of DNRC distribution shows a decreasing trend, while the probability curve shows an increase.
The bar at the abscissa of 16 also includes the number of blocks whose DNRCs are greater than 16.
Due to a large number of background blocks, the number of blocks with a zero DNRC value account for a significant proportion (about 96\%).
The probability of these blocks being motion blocks is merely about 1\%.
Therefore, we focus only on blocks with non-zero DNRCs.
%It is reasonable that the probability curve should show a small amount of fluctuation between DNRC 13 and 16, since the percentage of blocks with high DNRC values is too small, and the count is not sufficiently large to indicate the real probability.
%The probability that a block with a DNRC value of 8 is a motion block is already higher than 80\%.
Based on the tendency of the curve in Fig.~\ref{DNRC_Analysis}, we can conclude that a
larger DNRC block has a higher probability of being a motion block.

\subsection{Running Time}
Without fully decoding or any decryption, we can extract the DNRC feature directly from video bitstreams, which is very cost-effective in terms of time.
%Since our object detection and tracking processes are integrated and complementary, we recorded the time taken by the whole process and then calculated the processing speed by dividing by the number of frames.
%The results for each video sequence are shown in Table \ref{Runningtime}.
%We can see that the processing speed of our scheme exceeds 100 fps, meaning that it is highly efficient and
%very satisfy the requirements of real-time processing.
The average processing speed of our scheme for all videos reaches 107.92 fps, meaning that it is highly efficient and satisfies the requirements of real-time processing.
The main factor contributing to this high speed is that the DNRC feature can be efficiently extracted and processed with no complicated calculations of the pixel values, as in optical flow.

\subsection{Impact of Parameter}
\begin{figure}[!t]
	%\subfigure{
	\centering	
	\includegraphics[width=2.5in]{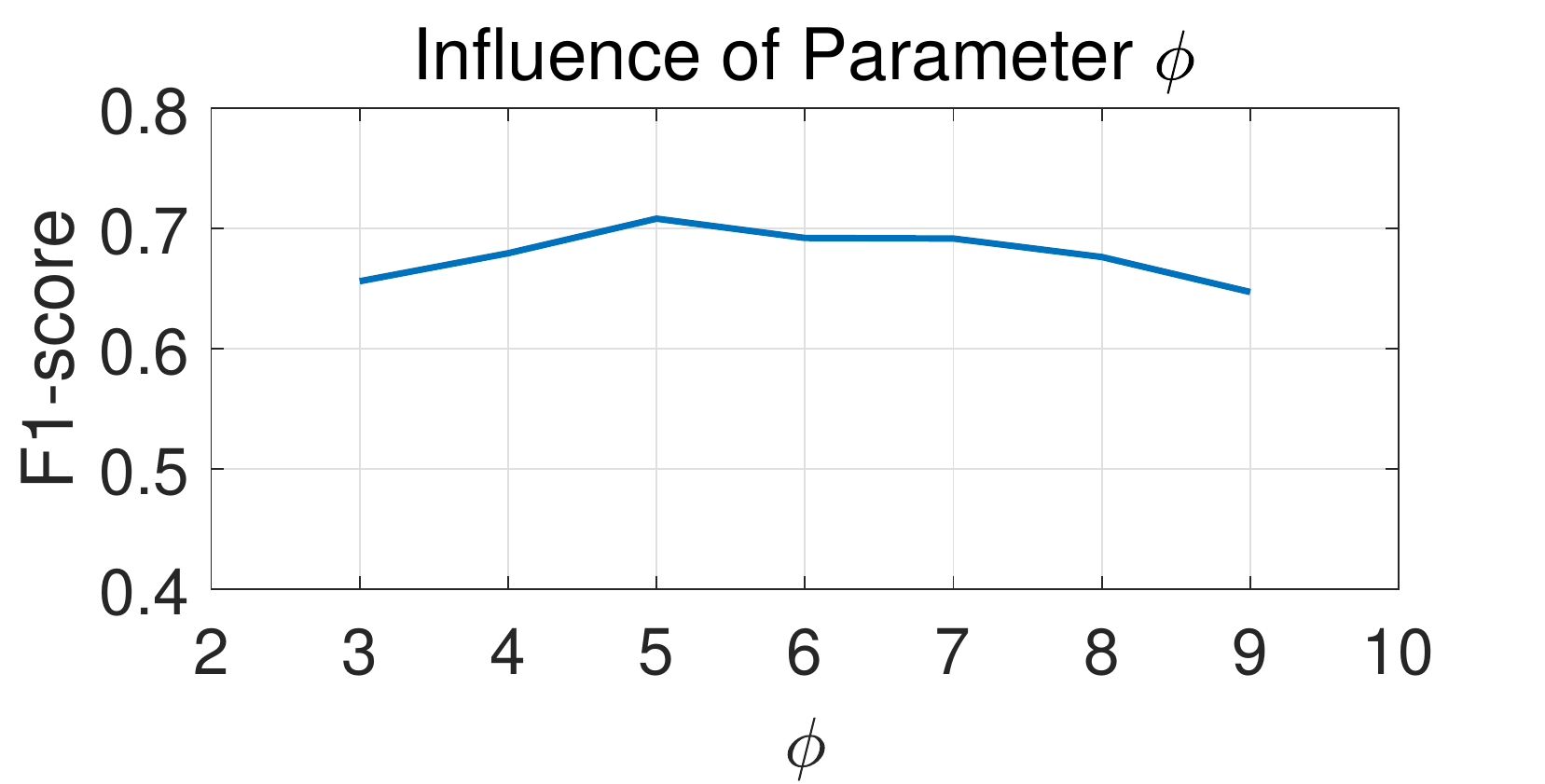}
	%}
	%	\subfigure{
	%		\centering
	%		\includegraphics[width=1.6in]{figure/alphaParameter.png}
	%	}
	\caption{The influence of the parameter $\mu$ on the average F1-score for all video sequences.}
	\label{parameter}
\end{figure}

Only one manual parameter is used in the proposed scheme, i.e., the time-domain filter parameter $\mu$.
%If there are sufficient training data for a specific surveillance scenario,
%we can obtain an optimal value of $\mu$ by using a grid search to achieve better performance.
We conducted experiments to investigate the sensitivity of the performance on the value of the manual parameter.
We vary the value of $\mu$ from 3 to 9 and then compute the average F1-score of object detection for all video sequences.
The results are shown in Fig. \ref{parameter}, from which we can see that the chosen of $\mu$ from common values has a small influence on our scheme's performance.

\begin{figure}[!t]
	\centering	
	\includegraphics[width=5in]{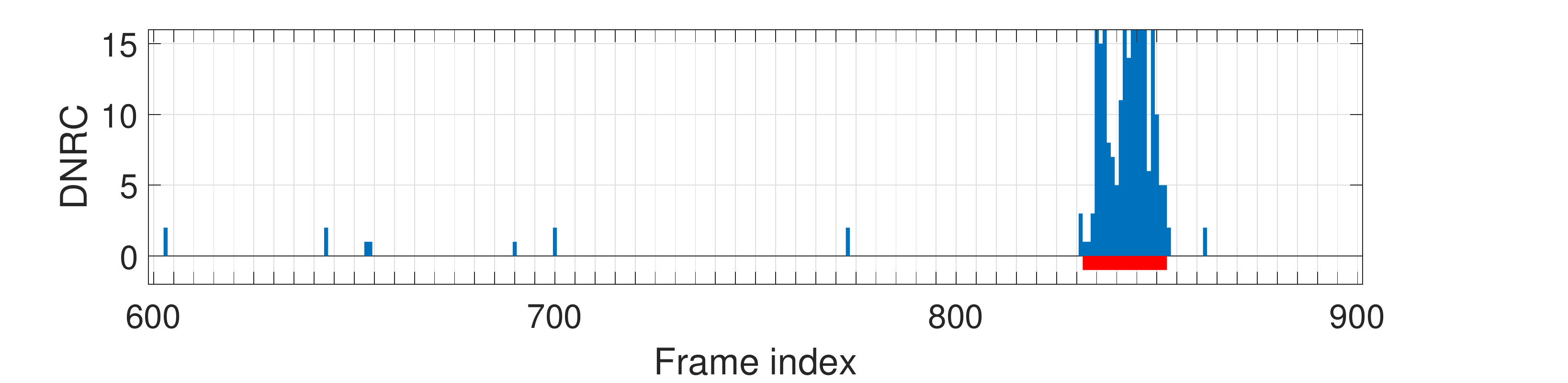}
	\caption{The distribution of noise spikes for a 4$\times$4 block in the video sequence of Pedestrians. The corresponding motion annotation is marked in red on the bottom. The spike noise is the blue bar without a red ground truth label. Each grid in X-axis denotes the interval of $\mu$ frames.}
	\label{spikenoise}
\end{figure}

To investigate the reality of the assumption, i.e., there is one noise spike approximately every $\mu$ frames. We have conducted experiments on the statistical distribution of noise spikes in the test video dataset. We show an example in Fig. \ref{spikenoise}. Every $\mu$ frames only contain one or two noise spikes. Our experimental results show that this assumption is reasonable in the scenario of video surveillance.

\subsection{Extension to other Video Coding Standards and Video
	Bitstream Encryption Schemes}\label{adaptation}
The proposed technique for object tracking is not specifically designed to rely on the H.264 video encryption scheme in this paper.
We can adapt the proposed scheme to other advanced video encoding frameworks, such as H.265 video compression, since residual data widely exist in these video compression standards.
The proposed detection and tracking scheme is also applicable to other video bitstream encryption methods.
We have conducted more experiments to investigate the effectiveness of our scheme on four different H.264/H.265 video encryptions~\citep{6725633,5733402,8119905,6589970}.
%Some of their encryption examples are given in Fig. \ref{bitstreamencrypt}.
As shown in Table \ref{adapttacking}, the experimental results demonstrate that the proposed motion detection and tracking technique also works for these tested video encryption schemes.
Note that the extracted DNRC features from two different encryption schemes under the same codec are the same.
Our motion and tracking scheme achieves the same performances under the same codec.

%\begin{figure}[]
%	\centering
%	\subfigure[Xu \citep{6725633}]{
%		\includegraphics[width=0.7in]{figure/xu.png}}
%	\subfigure[Shahid \citep{5733402}]{	
%		\includegraphics[width=0.7in]{figure/shahid264.png}}
%	\subfigure[Sallam \citep{8119905}]{
%		\includegraphics[width=0.7in]{figure/sallam.png}}
%	\subfigure[Shahid \citep{6589970}]{
%		\includegraphics[width=0.7in]{figure/shahid265.png}}
%	\caption{\textcolor{black}{Video bitstream encryption examples on the frame \#100 in the \emph{FOREMAN} video sequence.} }
%	\label{bitstreamencrypt}
%\end{figure}
\subsection{Analysis of Privacy Protection}
%We further conducted an experiment to investigate the privacy protection performance of our video encryption resisting to practical face detection and recognition algorithms. We adopted the most popular project of face recognition on Github \citep{ageitgeyface} in our experiment. The experimental results are shown in Fig. \ref{facedetection}. Given a known girl face in the \emph{akiyo} video sequence, we detect two human faces in the news video sequence and successfully identify the face of \emph{akiyo} in plaintext video. However, face detection and recognition algorithms fail to distinguish faces in the encrypted \emph{news} video. Therefore, we can learn that the employed video encryption can provide sufficient privacy guarantees for de-identification in practice.}
%
%\begin{figure}[!t]
%	\centering	
%	\subfigure[\emph{akiyo}]{
%		\includegraphics[width=1in]{figure/akiyo.jpg}}
%	\subfigure[original \emph{news}]{
%		\includegraphics[width=1in]{figure/news.jpg}}
%	\subfigure[encrypted \emph{news}]{
%		\includegraphics[width=1in]{figure/enccryptednews.jpg}}
%	\caption{\textcolor{black}{Examples of face detection and recognition results.}}
%	\label{facedetection}
%\end{figure}

We conducted experiments to investigate the performance of privacy protection with our video encryption under the threat models T1 and T3 \citep{9320277}.
T1 assumes the attacker has no information of any obscuration method
while
T3 assumes the attacker knows the exact type of the obscuration method and its hyperparameter.
%Following with \citep{9320277}, three types of face attack based on deep learning model are adopted in the experiments, including identification, verification, and reconstruction.
We make all experimental settings to be consistent with \citep{9320277}.
The used datasets in \citep{9320277} include FaceScrub dataset \citep{ng2014data} and LFW dataset \citep{Huang2007labeled}, which are encoded into
video bitstreams for our video encryption.
Three types of attacks are used for examining the robustness of the employed video encryption scheme, including face identification attack, face verification attack, and face reconstruction attack.
Two popular models of VGG-19~\citep{simonyan2014very} and ResNet-50~\citep{He_2016_CVPR} are adopted as the model backbone.
The performance of identification and verification attack are evaluated by Top 1 accuracy and AUC of ROC, respectively. Mean square error (MSE) between the clear and reconstructed image and identification accuracy of recovered images are adopted as the evaluation metrics for reconstruction attack.
The experimental results are given in Table~\ref{identification}.

The identification accuracies of our video encryption under T1 with two backbones are less than 0.009.
Under T3, the identification accuracies of the employed video encryption of two backbones are less than 0.18.
Compared with other obscured methods under T3,
the performance of the employed encryption scheme significantly outperforms the traditional obscuration methods and privacy-preserving image encryption methods,
and is closer to that of $k$-same based obscuration methods.
For the face verification attack under T3, our video encryption with VGG-19 achieves an AUC of
0.646, which are smaller than the original $k$-same method~\citep{10.1007/11767831_15} but greater than those of $k$-same-net~\citep{e20010060} and UP-GAN~\citep{hao2019utility}.
In general, the average AUC value of our video encryption are at the same level as those of the $k$-same based methods.
As for the face reconstruction attack,
%the MSE value of our video encryption is larger than all the methods in~\citep{9320277}.
%The accuracy of our encryption scheme is 0.001, which is smaller than all the methods in~\citep{9320277}.
our video encryption achieves the highest MSE and the lowest identification accuracy.
Thus, the employed encryption scheme performs better than the $k$-same based obscuration methods in resisting reconstruction attack.
According to the experimental results of the above three types of face attacks, the employed video encryption achieves a similar privacy level with the best methods in~\citep{9320277}, i.e., the k-same based obscuration methods.
Therefore, we can draw the conclusion that the employed video encryption scheme can achieve a considerable obscuration performance even under the strongest attack model T3.

%The video privacy is well protected in our motion detection and tracking system.

%The face attack results of identification and verification are given in Table~\ref{identification} and Table~\ref{verification}, respectively.
%	Under the threat model T3, the face identification accuracies of the employed video encryption using VGG-19 \citep{simonyan2015deep} and ResNet-50 \citep{He_2016_CVPR} are 0.172 and 0.179, respectively. The face verification AUC of our video encryption using VGG-19 and ResNet-50 are 0.646 and 0.625, respectively.
%For face reconstruction attack, our encryption scheme achieves a MSE of 0.427 and an identification accuracy of 0.001.
%Some examples of the reconstructed face are given in Fig. \ref{fake_face}.
%Compared with the reported results in \citep{9320277}, we can learn that the employed video encryption achieves a similar privacy level with the best methods in \citep{9320277}, i.e., the k-same \citep{10.1007/11767831_15} based obscuration methods.
%So, we can draw the conclusion that the employed video encryption scheme can achieve a considerable obscuration performance even under the strongest attack of threat model T3.

For privacy protection, it is not necessary to encrypt every bit of video stream.
For example, the $k$-same obscuration methods, which were reported to be able to provide secured privacy protection for face images in \citep{9320277}, are also not conventional encryption schemes.
According to the conventional video compression standards, e.g., H.264 and H.265,
%~\citep{1218189,6316136},
identifiable information such as human faces is relative to the intra-frame video coding.
In contrast, the motion information is connected to inter-frame video coding.
%Thus, the employed video encryption allows performing motion detection does not mean that the video encryption method cannot protect video privacy.
Thus, privacy protection and motion detection are not mutually exclusive in our application scenario of video surveillance.
Specifically, from the above experimental results,
we can learn that the employed video encryption has a similar face obscuration performance with the best methods in \citep{9320277}.

%according to the experimental results of face attack, we can learn the employed video encryption has a similar face obscuration performance with the best methods in \citep{9320277}.
%}

\begin{table}
	
	\caption{Comparison of our encryption with other methods reported in~\citep{9320277} on the performance of resisting three types of face attack. The method of \emph{Clear} means the result of the clear image. The methods marked by \dag, \ddag, and * are the traditional obscuration method, the k-same based method, and the privacy-preserving image encryption method, respectively.}
	\label{identification}
	\centering
	\renewcommand{\arraystretch}{1.1}
		\begin{tabular}{c|cccc|cccc|cc}
			\hline
			\multirow{3}{*}{Method} & \multicolumn{4}{c|}{Accuracy of Identification Attack}                            & \multicolumn{4}{c|}{AUC ROC of Verification Attack}                             & \multicolumn{2}{c}{\multirow{2}{*}{\begin{tabular}[c]{@{}c@{}}Reconstruction\\ Attack\end{tabular}}} \\ \cline{2-9}
			& \multicolumn{2}{c}{VGG19}       & \multicolumn{2}{c|}{ResNet50}    & \multicolumn{2}{c}{VGG19}      & \multicolumn{2}{c|}{ResNet50}   & \multicolumn{2}{c}{}                                                                                 \\ \cline{10-11}
			& T1             & T3             & T1             & T3             & T1             & T3             & T1             & T3             & MSE                                               & Acc                                              \\ \hline
			Clear                   & 0.838                  & 0.886                  & 0.890                  & 0.884                   & 0.983          & 0.983          & 0.981          & 0.981          & 0.000                                             & 0.849                                            \\
			Gaussian\dag                & 0.007                  & 0.811                  & 0.009                  & 0.798                   & 0.512          & 0.893          & 0.629          & 0.962          & 0.002                                             & 0.367                                            \\
			Median\dag                  & 0.011                  & 0.805                  & 0.014                  & 0.798                   & 0.539          & 0.877          & 0.592          & 0.933          & 0.007                                             & 0.102                                            \\
			Pixelation\dag              & 0.004                  & 0.373                  & \textbf{0.002}         & 0.323                   & 0.505          & 0.630          & 0.530          & 0.792          & 0.031                                             & 0.006                                            \\
			k-same\ddag                  & 0.012                  & \textbf{0.050}         & 0.012                  & \textbf{0.063}          & 0.573          & 0.695          & 0.580          & 0.768          & 0.029                                             & 0.005                                            \\
			k-same-net\ddag              & 0.091                  & 0.095                  & 0.081                  & 0.092                   & 0.505          & 0.497          & \textbf{0.493} & \textbf{0.492} & 0.064                                             & 0.018                                            \\
			UP-GAN\ddag                  & 0.091                  & 0.093                  & 0.082                  & 0.088                   & \textbf{0.500} & \textbf{0.494} & 0.499          & 0.497          & 0.059                                             & 0.003                                            \\
			P3*                      & \textbf{0.001}         & 0.678                  & \textbf{0.002}         & 0.579                   & 0.503          & 0.524          & 0.502          & 0.899          & 0.013                                             & 0.339                                            \\
			Scrambling*              & 0.002                  & 0.784                  & \textbf{0.002}         & 0.750                   & 0.544          & 0.951          & 0.549          & 0.928          & 0.018                                             & 0.042                                            \\
			Ours                    & 0.007                  & 0.172                  & 0.009                  & 0.179                   & 0.585          & 0.646          & 0.575          & 0.625          & \textbf{0.427}                                    & \textbf{0.001}                                   \\ \hline
		\end{tabular}

\end{table}
\begin{figure}[]
	\centering
	\subfigure{
		\includegraphics[width=0.55in]{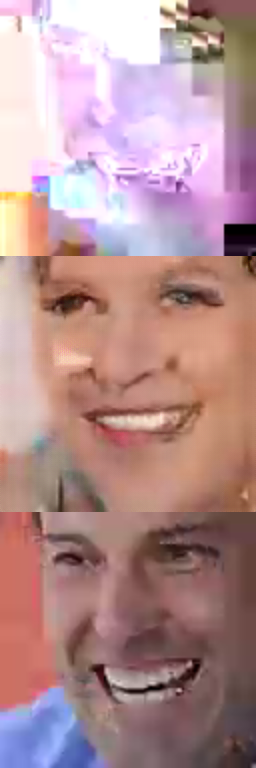}}\vspace{1pt}
	\subfigure{	
		\includegraphics[width=0.55in]{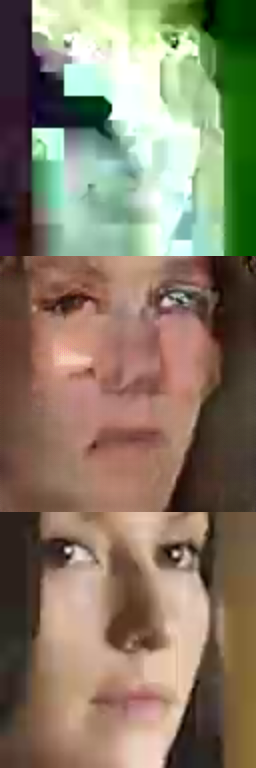}}\vspace{1pt}
	\subfigure{
		\includegraphics[width=0.55in]{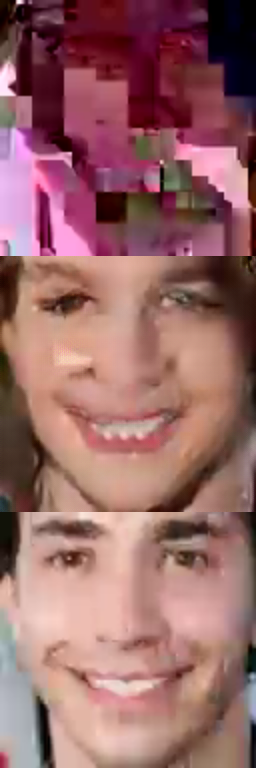}}\vspace{1pt}
	\subfigure{
		\includegraphics[width=0.55in]{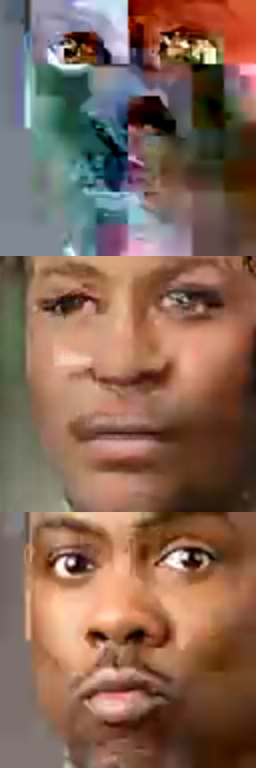}}\vspace{1pt}
	\subfigure{
		\includegraphics[width=0.55in]{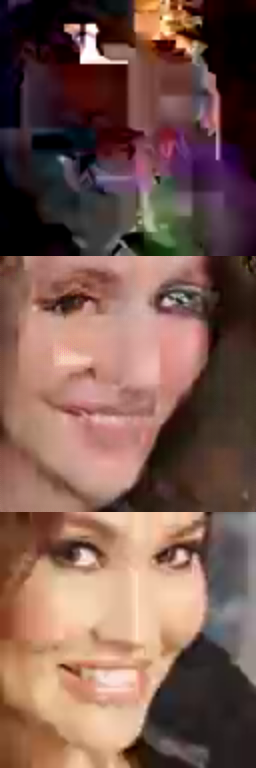}}\vspace{1pt}
	\subfigure{
		\includegraphics[width=0.55in]{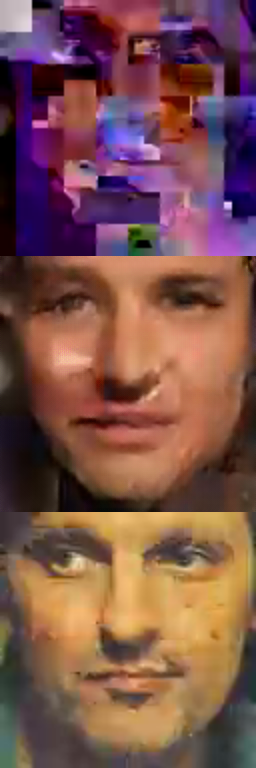}}\vspace{1pt}
	\subfigure{
		\includegraphics[width=0.55in]{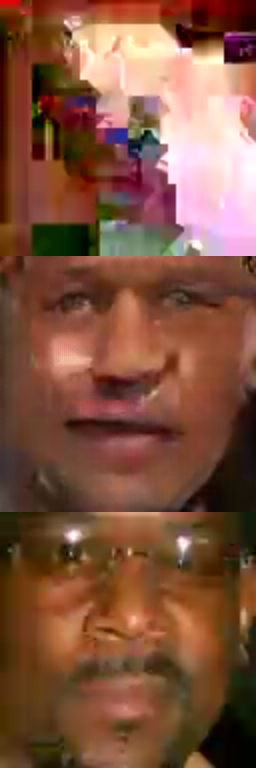}}
	\caption{Examples of the reconstruction attack. The top picture is the encrypted face, the middle is the reconstructed face, and the bottom is the unencrypted face.}
	\label{fake_face}
\end{figure}
\section{Conclusions}\label{conclusion}

In this paper, we propose a secure and robust motion detection and object tracking scheme based on a new feature, to tackle complex and dynamic scenes in real surveillance tasks.
The proposed feature DNRC is extracted from compressed video bitstream directly without decryption and full decompression.
By outsourcing the detection and tracking tasks to the cloud, we can obtain responses for
the moving object regions from the cloud that processes the encrypted and compressed videos.
%In our scheme, the privacy of the video data is protected against the cloud.
Meanwhile, the cloud can provide automatic services of effective small-sized object detection, robust and reliable detected object tracking in complex surveillance scenarios such as occlusion, camera moving, shadow, etc.
Our experimental results show that the proposed scheme achieves outstanding performance in moving object detection and tracking in the encrypted domain.
%Our scheme even shows competitive performance in the plaintext domain in comparison with some popular
%pixel-domain-based methods.
%An analysis of the running time illustrates that our algorithm keeps the computational complexity as low as
%possible in H.264/AVC compressed domain, and therefore meets the requirements for real-time operation.
%\textcolor{red}{The security analyses verify that our privacy-preserving system has sufficient privacy guarantee. Furthermore, the proposed algorithm has good compatibility both on encryption schemes and recent codecs.}
%and it was also shown that our approach has good efficiency and robustness.
%The proposed scheme also has broad flexibility, and can overcome the various challenges faced by motion detection and tracking systems.
The proposed algorithm has good compatibility both on encryption schemes and recent codecs and can overcome the various challenges of motion detection and tracking systems in real-time.

\bibliographystyle{unsrtnat}
\bibliography{IEEEabrv,references}  %%% Uncomment this line and comment out the ``thebibliography'' section below to use the external .bib file (using bibtex) .

%%% Uncomment this section and comment out the \bibliography{references} line above to use inline references.
% \begin{thebibliography}{1}

% 	\bibitem{kour2014real}
% 	George Kour and Raid Saabne.
% 	\newblock Real-time segmentation of on-line handwritten arabic script.
% 	\newblock In {\em Frontiers in Handwriting Recognition (ICFHR), 2014 14th
% 			International Conference on}, pages 417--422. IEEE, 2014.

% 	\bibitem{kour2014fast}
% 	George Kour and Raid Saabne.
% 	\newblock Fast classification of handwritten on-line arabic characters.
% 	\newblock In {\em Soft Computing and Pattern Recognition (SoCPaR), 2014 6th
% 			International Conference of}, pages 312--318. IEEE, 2014.

% 	\bibitem{hadash2018estimate}
% 	Guy Hadash, Einat Kermany, Boaz Carmeli, Ofer Lavi, George Kour, and Alon
% 	Jacovi.
% 	\newblock Estimate and replace: A novel approach to integrating deep neural
% 	networks with existing applications.
% 	\newblock {\em arXiv preprint arXiv:1804.09028}, 2018.

% \end{thebibliography}

\end{document}